\documentclass[preprint,12pt]{elsarticle}

\usepackage{multirow,color}
\usepackage{color}
\usepackage{colortbl}
\usepackage{adjustbox}
\usepackage{multirow,color}
\usepackage{hyperref}
\usepackage{graphicx}
\usepackage{forest}
\usepackage{tikz-qtree}
\usepackage{subcaption}
\usepackage{tabularx}
\usepackage{booktabs}
\usepackage{makecell}
\usepackage{pdflscape}
\usepackage{longtable}

\journal{Computer Science Review}

\begin{document}

\begin{frontmatter}



\title{Workflow as a Service Broker in Cloud Environment: A Systematic Mapping Study}

\author{Saeid Abrishami \corref{cor1}}
\ead{s-abrishami@um.ac.ir}
\cortext[cor1]{Corresponding Author}
\affiliation{
	\organization = {Department of Computer Engineering, Faculty of Engineering , Ferdowsi University of Mashhad},
	\city = {Mashhad},
	\country = {Iran}
}

\author{Faridreza Momtaz Zandi \fnref{fn1}}
\ead{faridreza.momtazzandi@studenti.polito.it}
\affiliation{%
	\organization={Departement of Control and Computer Engineering, Politecnico di Torino},
	\city = {Torino}, 
	\country = {Italy}
}

\author{Alireza Nourbakhsh \fnref{fn1}}
\ead{alireza.nourbakhsh@mail.um.ac.ir}
\fntext[fn1]{Both authors contributed equally to this research.}
\affiliation{%
	\organization={Department of Computer Engineering, Faculty of Engineering , Ferdowsi University of Mashhad},
	\city={Mashhad},
	\country={Iran}
}

%\authornote{corresponding  author}
%\email{s-abrishami@um.ac.ir}

%\email{faridreza.momtazzandi@studenti.polito.it}

%\authornotemark[2]
%\email{alireza.nourbakhsh@mail.um.ac.ir}

\begin{abstract}
Cloud computing has emerged as a promising platform for running scientific workflows across various domains. Scientists can take advantage of different cloud service models, such as serverful or serverless, to execute workflows based on their specific requirements, along with diverse pricing models like on-demand, reserved, or spot instances to reduce execution costs. However, the challenge of selecting appropriate resources and pricing models, coupled with the orchestration and scheduling of workflow tasks, creates significant complexity for users. To mitigate this burden, Workflow as a Service (WaaS) brokers have been introduced to facilitate workflow execution. In recent years, numerous studies have been published, either directly or indirectly related to this research area, highlighting the need for a comprehensive and systematic review of WaaS brokers to identify key trends and challenges in this field. In this paper, we conduct a Systematic Mapping Study (SMS) on WaaS brokers within cloud environments. The SMS employs a thorough 3-tier strategy (database search, backward snowballing, and forward snowballing) to answer five research questions. A total of 87 high-quality articles, published in 49 prestigious venues, are analyzed to derive a taxonomy based on the architecture of WaaS brokers. The articles are classified and surveyed according to this taxonomy, and future research directions for the design and implementation of WaaS brokers are explored. This study provides valuable insights for researchers and developers, helping them identify major trends and issues in the field of WaaS brokers.
\end{abstract}

\begin{highlights}
\item Conducts the first systematic mapping study on Workflow-as-a-Service brokers.
\item Develops a structured taxonomy for WaaS broker components and architectures.
\item Analyzes 87 primary studies to reveal prevailing trends and open research issues.
\item Outlines future research toward AI-driven, serverless, and hybrid quantum/classical brokers.
\end{highlights}

%% Keywords

\end{frontmatter}

\section{Introduction}
\label{sec:introduction}
Cloud computing has become one of the most transformative technologies of the modern era, reshaping industries and revolutionizing how computing resources are accessed and utilized. It enables customers to rent resources on demand and pay on a pay-as-you-go basis, eliminating the need for large upfront investments in infrastructure \cite{josepview2010view}. Both industry and academia have widely adopted this computational paradigm to meet their growing and diverse resource requirements \cite{versluis2021survey}. In particular, emerging paradigms in artificial intelligence (AI)—such as deep learning and large language models—benefit from the cloud’s ability to provide scalable access to resource-intensive infrastructure like GPUs. In response to evolving user demands, leading cloud service providers (CSPs) continue to innovate by offering new services aimed at improving usability, flexibility, and performance. As a result, beyond traditional offerings like Infrastructure as a Service (IaaS), which provides virtual machines (VMs), a range of higher-level services has emerged. These include Container as a Service (CaaS), Function as a Service (FaaS)—also known as serverless computing \cite{shafiei2022serverless}—and Quantum as a Service (QaaS), also referred to as Quantum Computing as a Service (QCaaS) \cite{kumari2025quantum}, each designed to abstract away infrastructure management and streamline application deployment.

\textit{Scientific workflows} are widely used to automate and execute complex scientific applications across diverse domains \cite{juve2013characterizing,abrishami2013deadline}. A scientific workflow defines a process aimed at achieving a scientific objective, typically represented as a set of tasks and their dependencies. To support the composition, planning, orchestration, and automated execution of such workflows, researchers commonly rely on \textit{Workflow Management Systems (WMSs)} \cite{suter2025terminology}—such as Pegasus \cite{deelman2019evolution}, Apache Airavata \cite{Airavata}, and Apache Airflow \cite{Airflow}. These applications usually require resource-rich computing environments, such as distributed infrastructures. Cloud computing offers an attractive solution, enabling scientists to exploit services such as IaaS, CaaS, and FaaS, together with pricing models like \textit{on-demand}, \textit{spot}, or \textit{reserved}, to execute workflows efficiently and cost-effectively. A central challenge, however, lies in selecting appropriate resources and pricing models and, more critically, orchestrating and scheduling workflow tasks across them. This is where \textit{cloud brokers} become essential. Brokers mediate between CSPs and customers, identifying suitable services while incorporating user preferences \cite{khorasani2024cloud}. They also mitigate the \textit{vendor lock-in problem}, i.e., dependence on a single CSP’s services, which may otherwise lead to monopolistic scenarios. 

Workflow as a Service (WaaS) represents a class of cloud brokers that accept users’ workflows together with their specified Quality of Service (QoS) requirements, such as \textit{deadline} and \textit{budget}, and execute them by provisioning and allocating suitable resources from one or more CSPs. While the broker seeks to maximize its own benefits, it must also provide sufficient incentives for users to outsource their workflows, most notably by reducing execution costs. To accomplish this, the broker relies on two primary components: the \textit{resource provisioner}, which is responsible for selecting resources across different service and pricing models, and the \textit{task scheduler}, which maps workflow tasks onto the provisioned resources. By executing multiple workflows concurrently, the scheduler can improve overall resource utilization compared to scheduling workflows independently. This makes scheduling in WaaS brokers a distinctive challenge, commonly referred to as the \textit{Multiple Workflow (or Multi-Workflow) Scheduling Problem (MWSP)}.

A structured review of WaaS brokers is crucial for uncovering trends, challenges, and gaps in this domain. To achieve such coverage, systematic methodologies are typically adopted, most notably the \textit{Systematic Literature Review (SLR)} \cite{kitchenham2009systematic} and the \textit{Systematic Mapping Study (SMS)} \cite{peterson2015guidelines}. While both approaches share a rigorous procedure for locating, filtering, and selecting studies, their objectives are fundamentally different. An SLR is a focused, evidence-oriented method designed to synthesize and compare existing research in depth, often including a critical evaluation of methods, metrics, and performance results under similar conditions. By contrast, an SMS does not aim to compare or rank methods. Instead, its role is broader and more exploratory: it maps the landscape of existing work, categorizes studies by dimensions such as method, contribution type, or evaluation strategy, and analyzes \textit{descriptive statistics} such as publication years, venues, and frequency of techniques. The emphasis is on identifying research trends and highlighting underexplored areas, thereby revealing gaps in the literature. In this sense, SMS research questions are intentionally general, seeking to provide a \textit{big picture} of the field, whereas SLRs drill down into fine-grained, evidence-based comparisons. An SMS often lays the groundwork for one or more SLRs by highlighting sub-areas that merit detailed comparative study.

To the best of our knowledge, no prior SMS has been conducted specifically on WaaS brokers. While a few studies address the broader domain of cloud brokerage \cite{li2022survey,khorasani2024cloud}, they do not specifically examine WaaS brokers. However, given that the scheduler is a key component of a WaaS broker, surveys on \textit{workflow scheduling} in cloud environments are considered related works. Several studies have surveyed cloud workflow scheduling in general \cite{versluis2021survey,adhikari2019survey,kumar2019comprehensive,masdari2016towards}, but while these articles touch on the MWSP, they do not comprehensively address it. Recently, Saeeidzade and Ashtiani \cite{saeedizade2024scientific} published a survey on workflow scheduling in cloud environments, which includes a subsection on scheduling in WaaS brokers. However, this survey does not comprehensively explore the domain. The most relevant work to this SMS is by Hilman et al. \cite{hilman2020multiple}, which provides a detailed survey of MWSP in multi-tenant distributed systems such as clusters, grids, and clouds. While this survey is comprehensive, it has several limitations.  First, its taxonomy encompasses all distributed systems, including aspects irrelevant to the cloud environment. Second, it was published in 2019 and reviews only 31 articles, whereas numerous new studies have emerged in the past five years, particularly in the field of WaaS brokers (as shown in Section \ref{sec:methodology}). This highlights the need for an updated survey. Finally, new technologies, such as serverless and quantum computing, have gained significant attention since the publication of Hilman et al.'s survey, further underscoring the necessity for a new review focused on recent developments in this area.  

We have conducted an SMS \cite{kitchenham2011using, peterson2015guidelines} in the field of WaaS brokers within the cloud environment. To identify studies related to this area, we designed queries based on key terms relevant to the field. A three-tier search strategy was employed, consisting of \textit{database search}, \textit{backward snowballing}, and \textit{forward snowballing} (see Section \ref{sec:methodology}). By integrating these three search strategies, 87 papers from 49 different venues (journals and conferences) were selected based on specific inclusion and exclusion criteria. These papers were then analyzed to answer 5 research questions, serving as a valuable resource for research teams and developers in the field. All included papers and the corresponding extracted data used to address the research questions are provided in Appendix B, as an Excel supplementary file available at \url{https://github.com/SaeidAbrishami/WaaSBroker}. 

The remainder of this article is organized as follows: Section \ref{sec:architecture} provides a detailed introduction to the architecture of WaaS brokers. Section \ref{sec:methodology} elaborates on the research methodology and provides a brief report of the review. Section \ref{sec:taxonomy} proposes a taxonomy for this field and surveys existing solutions based on this taxonomy. The future directions of WaaS brokers are outlined in Section \ref{sec:future}, and Section \ref{sec:conclusion} concludes the article. Furthermore, Appendix A discusses real-world adoption and industrial implementations of WaaS brokers.

\section{Architecture of a WaaS broker}
\begin{figure}
	\includegraphics[scale=0.9]{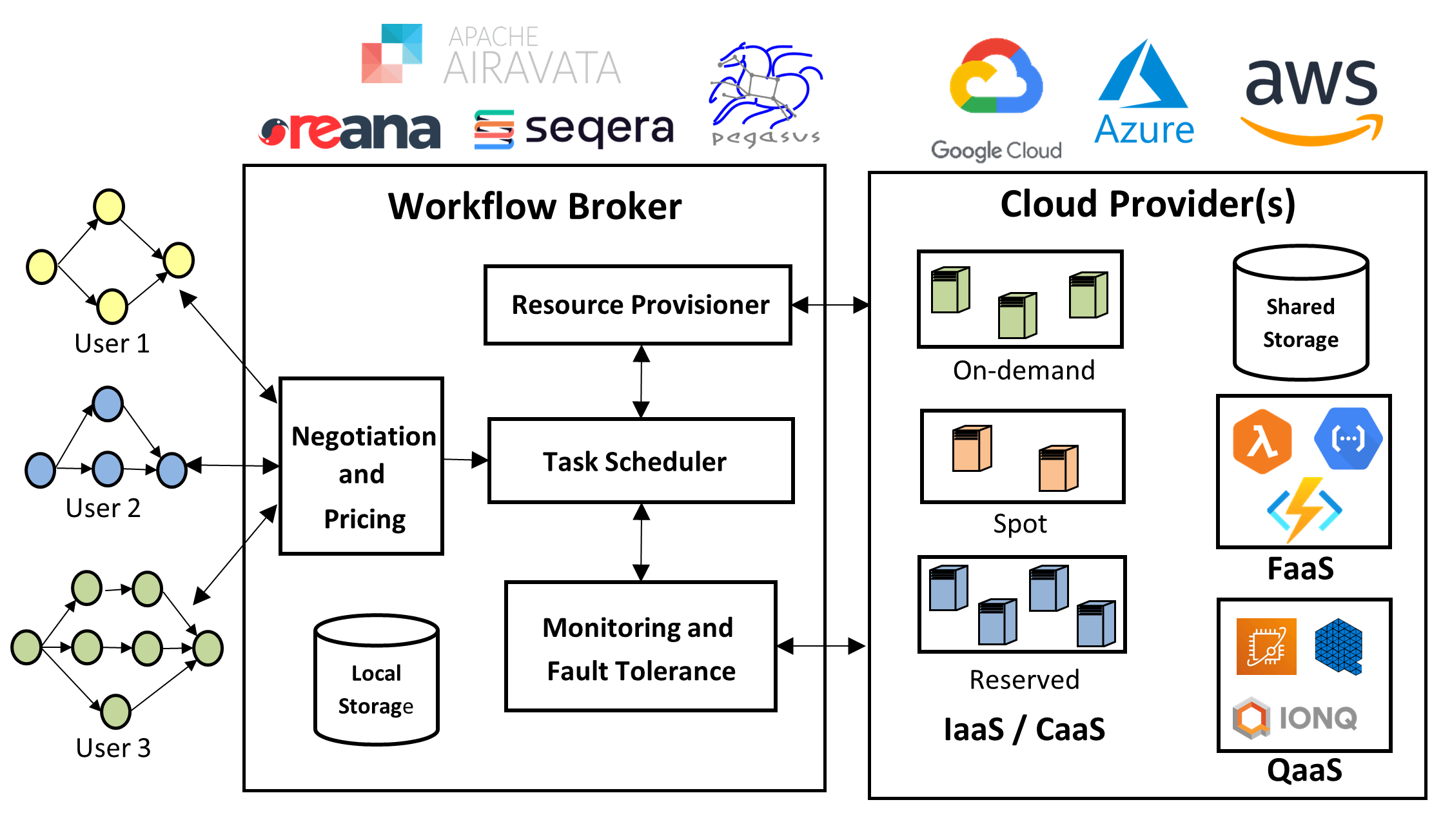}
	\caption{Architecture of a Workflow as a Service broker.}
	\label{fig:WaaS}
\end{figure}

Figure \ref{fig:WaaS} illustrates the architecture of a typical WaaS broker, as proposed based on both academic literature \cite{saeedizade2024scientific, adhikari2019survey, hilman2020multiple, 1, 2, 38, 49, 32, 57, 8, 53, 59} and real-world implementations \cite{deelman2019evolution, Airavata}. As shown, there are three main actors in the system: \textit{users}, \textit{cloud service provider(s)}, and the \textit{WaaS broker}. The process begins when users send one or more workflows to the WaaS broker. A workflow is usually represented as a \textit{Directed Acyclic Graph (DAG)}, where nodes denote tasks and edges capture data or control dependencies between them. This dependency indicates that a child task can only begin execution once its parent task(s) has completed and the child task has received the necessary output data. Additionally, users can often specify QoS parameters, such as deadlines and maximum budgets, for executing the workflow through a Service Level Agreement (SLA). In some cases, the broker may agree to a (monetary) penalty if the SLA is violated.  

CSPs—particularly the three major ones: Amazon Web Services (AWS), Google Cloud Platform (GCP), and Microsoft Azure—offer the essential services and resources required to execute users' workflows. A broker can provision these resources from a single CSP, multiple CSPs, or a hybrid cloud environment that combines private (on-premise) and public cloud infrastructures (see Section \ref{sec:deployment}). Currently, four widely used cloud services support the execution of scientific workflows: IaaS, CaaS, FaaS, and QaaS. Traditional IaaS provides users with VMs featuring various hardware and software configurations. However, the relatively long provisioning time of heavyweight VMs makes IaaS less suitable for small- to medium-sized tasks in traditional scheduling models, where each VM is typically assigned to a single task.

This limitation is addressed by CaaS, a newer model that offers lightweight virtualized environments known as containers. Containers have gained popularity for managing cloud applications due to their efficiency and flexibility \cite{pahl2019cloud}. Each container operates in an isolated environment with dedicated CPU, memory, block I/O, and network resources, enabling multiple tasks to run concurrently on a single VM. This allows task schedulers to place several small or medium-sized tasks on the same VM, achieving higher resource utilization compared to the IaaS model. Popular container services include Amazon Elastic Container Service (ECS), Azure Container Instances (ACI), and Google Kubernetes Engine (GKE).

\label{sec:architecture}

Another recent service model offered by CSPs is serverless computing, typically realized through FaaS in combination with Backend-as-a-Service (BaaS). FaaS allows developers to write stateless, event-driven functions, enabling them to focus solely on application logic without managing the underlying infrastructure (e.g., VMs or containers), which is fully managed by the CSP \cite{shafiei2022serverless}. Popular FaaS offerings include AWS Lambda, Azure Functions, and Google Cloud Functions. FaaS simplifies workflow execution by allowing tasks to be defined as individual functions, with resource provisioning and scaling handled automatically by the CSP. However, certain inherent characteristics of FaaS platforms introduce challenges. For instance, the stateless nature of functions requires the scheduler to rely on external storage services to store and exchange intermediate data between tasks. To address this, object storage solutions such as AWS S3, Azure Blob Storage, and Google Cloud Storage (GCS) are commonly used. When file-based access is required, shared file storage services—such as AWS Elastic File System (EFS), Azure Files, and Google Cloud Filestore—can be employed.

Recently, Quantum Computing (QC) has emerged as a revolutionary technology with the potential to transform numerous scientific domains by solving computational problems that are intractable for even the most advanced classical computers. Applications such as quantum chemistry, combinatorial optimization, and machine learning stand to benefit significantly from quantum speedups \cite{gemeinhardt2023quantum}. %Quantum computers operate using Quantum Processing Units (QPUs), which rely on qubits instead of classical bits. Unlike classical bits that represent either 0 or 1, qubits can exist in a superposition of both states simultaneously, enabling powerful parallelism. Today’s quantum machines—referred to as Noisy Intermediate-Scale Quantum (NISQ) devices—feature a limited number of qubits, are prone to noise, and have short qubit coherence times. Despite these limitations, promising quantum algorithms and applications have already been developed for NISQ systems.
Due to the complexity and high cost of operating quantum hardware—along with strict environmental requirements—on-premise deployment is not practical for most users. As a result, quantum computing is typically offered through cloud-based platforms, a model known as QaaS \cite{kumari2025quantum}. Notable QaaS platforms include Amazon Braket \cite{AWSBraket}, Azure Quantum \cite{AzureQuantum}, Google Quantum AI \cite{GoogleQuantumAI}, IBM Quantum Platform \cite{IBMPlatform}, and IonQ \cite{IonQ}. These services give scientists and developers access to real quantum hardware and simulators, enabling experimentation and algorithm development without needing direct access to physical quantum computers. 

%\begin{figure}
%	\includegraphics[scale=0.7]{figs/HybridWorkflow.jpg}
%	\caption{Hybrid quantum/classical workflow for variational quantum eigensolvers \cite{cranganore2024paving}.}
%	\label{fig:hybridWorkflow}
%\end{figure}

Contemporary quantum applications are predominantly hybrid, involving both quantum and classical components. Typically, quantum tasks are executed on Quantum Processing Units (QPUs), while classical computing resources are used for pre-processing, post-processing, and coordination tasks \cite{cranganore2024paving}. A well-known class of such applications is Variational Quantum Algorithms (VQAs) \cite{gemeinhardt2023quantum, cranganore2022molecular}, which follow an iterative quantum–classical interaction model. In VQAs, a parameterized quantum circuit is evaluated by a quantum computer, and the resulting cost function is optimized using a classical algorithm such as gradient descent. This results in a feedback loop: the quantum device computes the current output, the classical optimizer updates the parameters based on the cost, and the updated parameters are passed back for re-execution—repeating until a stopping condition is met. These applications can be effectively modeled as \textit{hybrid scientific workflows}, comprising a sequence of interconnected classical and quantum tasks \cite{cranganore2024paving, cranganore2022molecular}. 
%Figure \ref{fig:hybridWorkflow} illustrates an example of a VQA-based workflow for Variational Quantum Eigensolvers (VQEs) \cite{cranganore2024paving}. 
Executing such workflows requires seamless integration between High-Performance Computing (HPC) systems and QC environments \cite{elsharkway2025integration, beck2024integrating, shehata2026bridging}. The provisioning and scheduling of hybrid workflows introduce a new set of challenges, exceeding those found in purely classical systems. Addressing these challenges is essential for the development of robust and efficient WaaS brokers capable of supporting hybrid quantum-classical workloads \cite{cranganore2024paving}.

Another important aspect of the services of the CSPs is the pricing model. \textit{On-demand} is the oldest and most well-known pricing model: the CSP offers various services at pre-determined prices, and the user can lease resources on a pay-as-you-go basis. Historically, users were charged by the hour, but now, major CSPs charge users per second. In contrast, the \textit{reserved} pricing model allows the user to lease resources for an extended period (typically one year or more) at a significant discount compared to on-demand resources. Finally, the \textit{spot} pricing model enables users to access the CSP's excess computing capacity at a substantial discount. However, these resources are highly unreliable, as the CSP can reclaim them with as little as two minutes' notice. Therefore, the user must plan accordingly to manage this uncertainty \cite{lin2022methods}. The final pricing model pertains to the FaaS model, in which the user is typically billed based on the number of function invocations, the time it takes for execution, and the computational resources allocated to the function instance, particularly memory. 

Figure \ref{fig:WaaS} shows that the WaaS broker consists of four components: \textit{negotiation and pricing}, \textit{task scheduler}, \textit{resource provisioner}, and \textit{monitoring and fault tolerance}. The \textit{negotiation and pricing} component interacts with the user. It receives the user's workflow along with its required QoS and determines if it can meet the requirements. If the answer is yes, it specifies the price for running the workflow. It is reasonable for the price to be lower than what the user would pay by directly leasing resources from the CSP. However, the WaaS broker must also generate a profit. This can be achieved by efficiently using leased resources, sharing them between different workflows, and utilizing pricing models such as reserved instances to reduce the cost of workflow execution.

The \textit{task scheduler} assigns workflow tasks to leased resources, aiming to optimize objectives such as \textit{makespan} or \textit{execution cost} while meeting QoS requirements. To ensure efficient resource usage, it considers tasks from all workflows jointly rather than scheduling each workflow in isolation. As a result, the scheduler directly addresses the MWSP. The \textit{resource provisioner} complements the scheduler by leasing resources as needed. It decides when, how many, and which types of resources (e.g., cores, memory) to acquire, as well as when to release them to avoid unnecessary cost. This component is irrelevant in FaaS platforms, where resource management is handled directly by the CSP.

The fourth component, the \textit{monitoring and fault tolerance}, ensures correct task execution and tracks the status of leased resources. In case of failures, it initiates recovery actions such as task re-execution—especially critical in spot pricing models, where resources are unreliable and may be reclaimed at short notice. Finally, the broker’s local storage supports workflow execution by maintaining input files (or container images) and metadata. This includes \textit{provenance data} for reproducibility and \textit{monitoring data} such as execution logs and resource usage records \cite{suter2025terminology}.
 
\section{Research Methodology}
\label{sec:methodology}
We have conducted an SMS on WaaS broker according to the guidelines of \cite{peterson2015guidelines,kitchenham2009systematic}. The used SMS methodology has four phases \cite{khorasani2024cloud}, which are elaborated in the following.

\subsection{Specifying the Scope and Research Questions}
Research questions (RQs) are the foundation of a systematic study. In fact, an SMS is an effort to answer RQs. In this study, we have defined 5 research questions as follows:

\textbf{RQ1:} \textit{How active is the field of WaaS brokers?} This question has been answered in Section \ref{sec:analyze}.

\textbf{RQ2:} \textit{What is the taxonomy of designing a WaaS broker, and how can related work be classified in this taxonomy?} This question has been answered in Section \ref{sec:taxonomy}.

\textbf{RQ3:} \textit{Which techniques have been used for solving the MWSP and to what extent?} This question has been answered in Section \ref{sec:techniques}.

\textbf{RQ4:} \textit{Where are the research articles published?} This question has been answered in Section \ref{sec:analyze}.

\textbf{RQ5:} \textit{What are the open issues in the field of WaaS brokers?} This question has been answered in Section \ref{sec:future}.

Our review spans 21 years, from the advent of cloud computing in 2005 to the first half of 2025. It considers studies that explicitly design a WaaS platform for cloud environments or propose algorithms for the MWSP from the client-side perspective.

\subsection{Planning the Search Process}
\label{sec:query}
The first step is specifying search strategies. In this review, we used a three-tier search strategy: \textit{database search}, \textit{backward snowballing}, and \textit{forward snowballing}. Database search is a common method in literature surveys, where the researcher searches well-known online databases using \textit{search strings}. To conduct the database search, we used \textit{IEEE Xplore}, \textit{ScienceDirect}, \textit{SpringerLink}, \textit{ACM Digital Library}, and \textit{Wiley Online Library}. After reviewing some relevant papers and through a trial-and-error procedure by the authors, the following three queries were used as \textit{search strings}:

\textbf{Q1:} ("multi-workflow" OR "multi Workflow" OR "multiple workflow" OR "workflow ensemble") AND cloud AND scheduling

\textbf{Q2:} workflow AND broker AND cloud AND scheduling 

\textbf{Q3:} ("workflow as a service" OR "WfaaS" OR "WaaS" OR "workflow-as-a-service") AND cloud

These search strings were applied to the \textit{Title}, \textit{Abstract}, and \textit{Keywords} sections of the papers to retrieve all relevant papers. All extracted papers and their venues were evaluated against predefined criteria, and the papers meeting those criteria were added to the set of \textit{included papers}, which was then sent to the snowballing step.

\textit{Snowballing} is done in two steps: \textit{backward} and \textit{forward}. \textit{Backward snowballing} involves searching the reference list of an article to find new related articles. \textit{Forward snowballing} involves searching among the citing papers of an article to find new related articles. To find the citing papers for an article, we used \textit{Google Scholar}. The newly extracted articles underwent the evaluation process, and those added to the included papers were sent back to the snowballing step, continuing until no new articles were added to the set of included papers.

\subsection{Planning the Study Selection Process}
The first step of this phase is determining the criteria for venue inclusion/exclusion to ensure that high-quality venues are selected. For journals, those indexed in the Journal Citation Reports (JCR) are included. For conferences, those indexed in the ICORE conference ranking \cite{ICORE} are added to the list of included conferences. All extracted studies that are not published in the included venues are discarded from further consideration.

The next step is to define the study selection strategy. Relevance was first assessed by screening each paper’s title, abstract, and keywords. When these elements did not provide sufficient information, the full text was reviewed. To be included, a study had to explicitly design a \textit{workflow as a service} or propose a \textit{multiple workflow scheduling algorithm} for the \textit{cloud environment}. Studies focusing on single-workflow scheduling or targeting other distributed environments—such as clusters, grids, edge, or fog—were excluded (e.g., \cite{chen2025multi}, which addresses a hybrid cloud–edge architecture). We also restricted the scope to articles that approach the MWSP from the client-side (or broker-side), excluding those that address it from the provider-side, such as scheduling tasks on physical machines (e.g., \cite{wang2023adaptive, wang2024electricity, fan2023energy, wang2025electricity}). Finally, studies limited to specific workflow types, such as simple pipelines rather than general workflows modeled as DAGs, were excluded (e.g., \cite{sun2025genetic, shen2019dynamic}).

\subsection{Conducting the SMS and Analyzing Results}
\label{sec:analyze}
\begin{figure}[t] % "[t!]" placement specifier just for this example
	\begin{subfigure}{0.6\linewidth}
	\includegraphics[width=\linewidth]{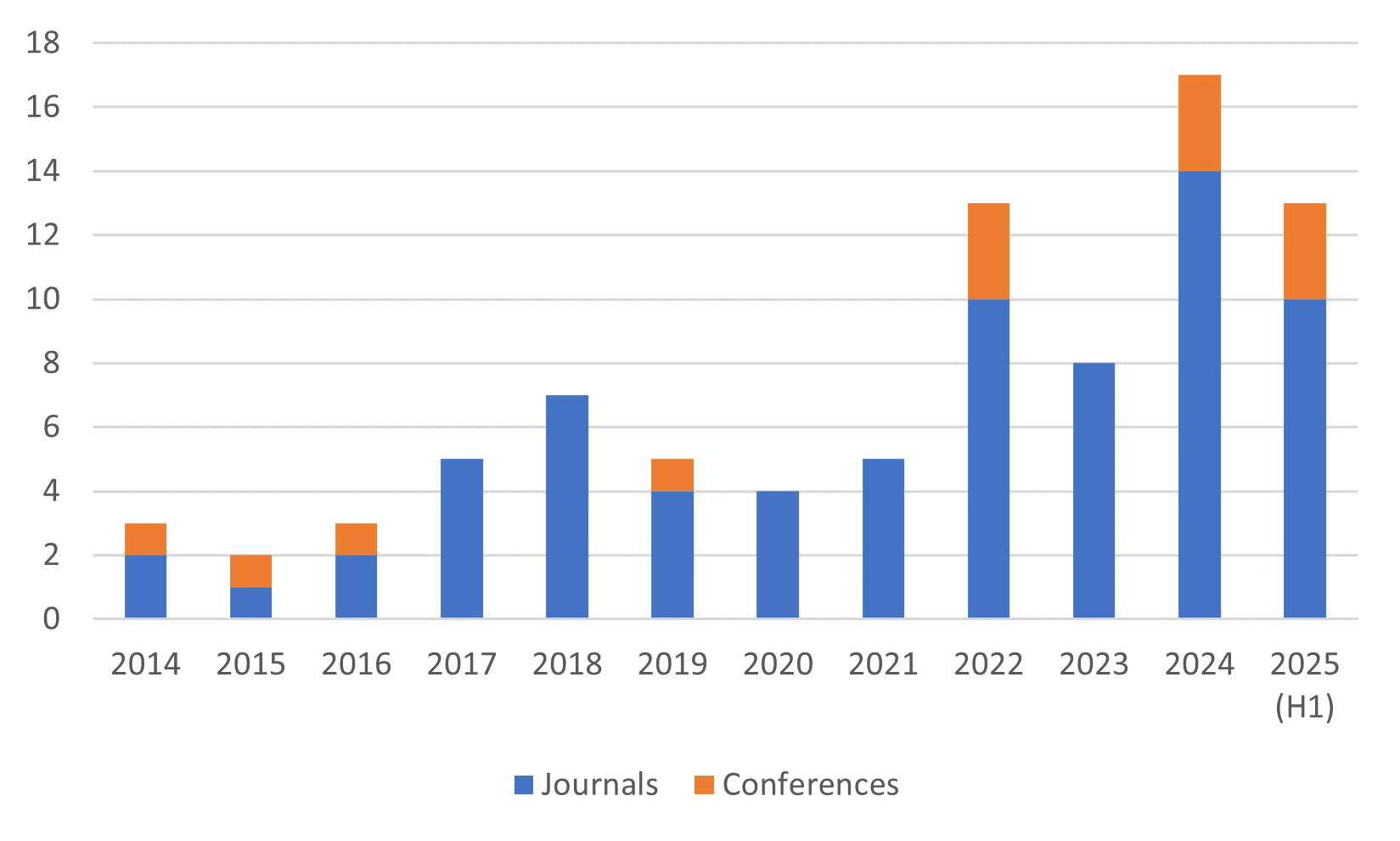}
	\caption{Per year}\label{fig:years}
	\end{subfigure}\hspace*{\fill}
	\begin{subfigure}{0.4\linewidth}
	\includegraphics[width=\linewidth]{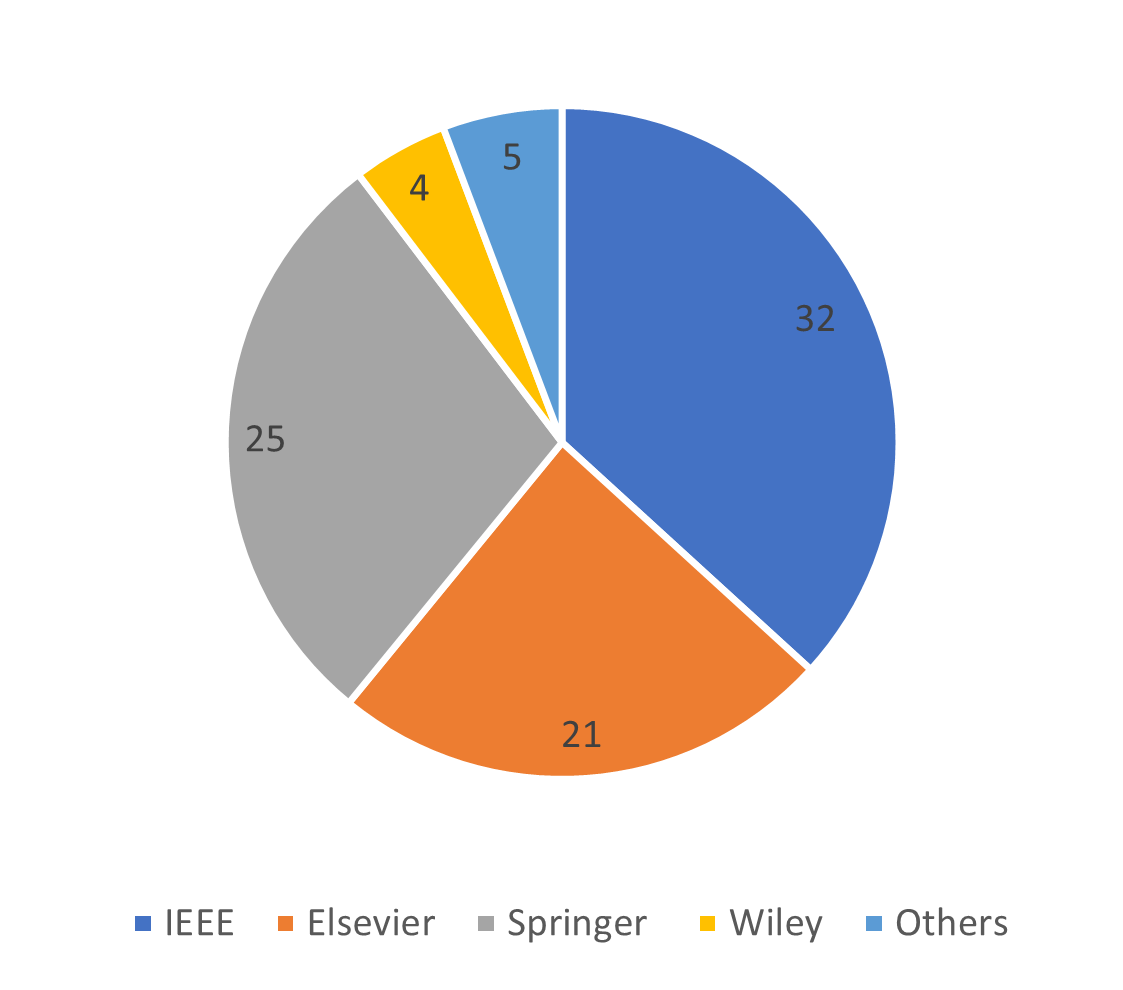}
	\caption{Per publisher}\label{fig:publishers}
	\end{subfigure}
	\caption{Number of included papers} \label{fig:published}
\end{figure}

We conducted the SMS following established strategies, identifying 49 venues (34 journals and 15 conferences) and 87 studies, all detailed in Appendix B. Figure \ref{fig:published} presents a brief overview of the studies included per year and per publisher. As shown in Figure \ref{fig:years}, the number of published papers has followed an upward trend in recent years, peaking in 2024 (with 2025 covering only the first half). The figure starts from 2014, as only two papers were published earlier (2009 and 2011). Figure \ref{fig:publishers} shows IEEE leads in publications, followed by Springer and Elsevier. The top three venues are \textit{The Journal of Supercomputing} with 8, \textit{Future Generation Computer Systems} with 7, and \textit{IEEE Transactions on Services Computing} with 6 published papers.

The final step involves analyzing and classifying the data to address the research questions of the SMS. The primary objectives of this SMS are to identify the taxonomy of the WaaS broker and to classify the included papers within this taxonomy. After a thorough review of the included papers, and based on the proposed architecture of the WaaS broker in Figure \ref{fig:WaaS}, a taxonomy was developed for each of the four components of a WaaS broker. All included papers were carefully analyzed by two authors and classified according to the taxonomy. Any disagreements were reviewed and resolved by all authors. The results of the analysis are presented in Section \ref{sec:taxonomy}.

\section{Taxonomy and survey}
\label{sec:taxonomy}
In this section, we elaborate on the proposed taxonomy and then survey the related work accordingly. The taxonomy is structured around the key components of a WaaS broker, which are discussed in the following four sub-sections, each dedicated to a specific component. To conserve space in this SMS, all included articles, along with their classifications under the proposed taxonomy, are provided in Appendix B as an Excel file. This format enables readers to conveniently search the articles and generate customized statistics and charts. In addition, Appendix A reviews real-world adoption and industrial implementations of WaaS brokers.

\subsection{Negotiation and Pricing Taxonomy}
\begin{figure}
    \centering
 
	\begin{forest} 
		for tree={
				font=\scriptsize, 
	    		edge path={\noexpand\path[\forestoption{edge}] (\forestOve{\forestove{@parent}}{name}.parent anchor) -- +(0,-10pt)-| (\forestove{name}.child anchor)\forestoption{edge label};}
		}
		[Negotiation and Pricing
			[QoS Requirements [Multi-criteria] [Constrained]]
			[Workflow Arrival [Online] [Batch] [Ensemble] [Periodic]]
			[Negotiation [Static]	[Dynamic]	]
			[Pricing [Pre-pricing] [Post-pricing]]
		]
	\end{forest}

    \caption{Negotiation and pricing taxonomy}
    \label{fig:negotiation}
\end{figure}

Figure \ref{fig:negotiation} presents the proposed taxonomy for the negotiation and pricing components, which is divided into four categories. Below, we provide a detailed explanation of each category.

\subsubsection{QoS Requirements}
Users have diverse requirements concerning QoS criteria, such as execution time, cost, privacy, reliability, and others (see Section \ref{sec:criteria}). These requirements can generally be divided into two categories. The first category is \textit{multi-criteria}, where the user expects multiple objectives to be optimized simultaneously. In this case, the broker is faced with a multi-objective optimization problem, which can be solved either by identifying Pareto-optimal solutions or by optimizing a weighted sum of the objectives. A large portion of studies in this category have concentrated on minimizing both the makespan and the execution cost of workflows \cite{7, 10, 33, 47}. Other studies have extended the optimization to include more objectives simultaneously, e.g., makespan, cost, and energy consumption \cite{32}, or makespan, cost, and fairness \cite{36}. 

\begin{figure}[t] % "[t!]" placement specifier just for this example
	\begin{subfigure}{0.4\linewidth}
	\includegraphics[width=\linewidth]{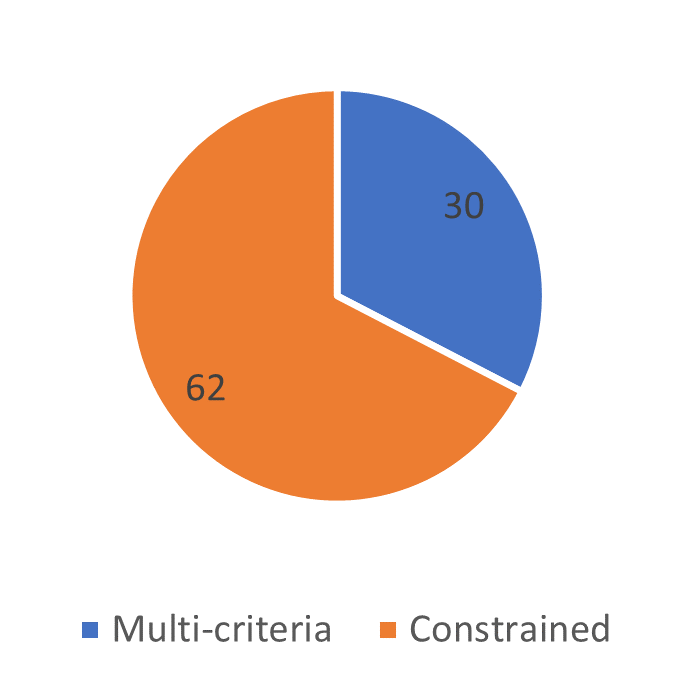}
	\caption{QoS Requirements}\label{fig:qos}
	\end{subfigure}\hspace*{\fill}
	\begin{subfigure}{0.4\linewidth}
	\includegraphics[width=\linewidth]{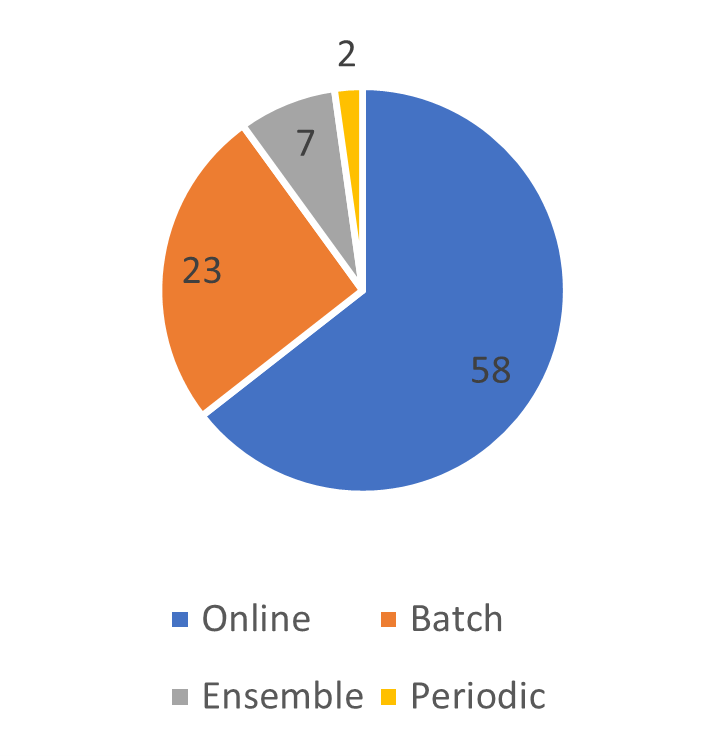}
	\caption{Workflow Arrival}\label{fig:arrival}
	\end{subfigure}
	\caption{Number of articles on QoS Requirements and Workflow Arrival.} \label{fig:qosarrival}
\end{figure}

On the other hand, the \textit{constrained} class focuses on optimizing one criterion while treating the other criteria as constraints. Two common constraints in this class are \textit{deadline-constrained} and \textit{budget-constrained} requirements. In a deadline-constrained scenario, the broker aims to meet the user-defined deadline while optimizing another criterion, typically minimizing the cost of workflow execution \cite{1, 19}. In a budget-constrained scenario, the broker seeks to keep the execution cost within the user-defined budget while optimizing other criteria, usually minimizing the workflow execution time \cite{48, 50}. Some studies address both deadline and budget constraints simultaneously \cite{27, 28, 15}. Finally, there are also studies that optimize multiple objectives while ensuring that certain constraints are met. For example, Karmakar et al. \cite{58} optimized resource utilization and rental costs while ensuring that the workflows met their deadlines. Figure \ref{fig:qos} indicates that the majority of previous studies adopted constrained QoS requirements.

\subsubsection{Workflow Arrival}
Workflow arrival patterns can be categorized into four classes. In the \textit{online} (or random) arrival pattern, workflows arrive at random times, with no predefined schedule \cite{4, 8, 11, 24}. As a result, the broker must be prepared to handle workflow arrivals at any time. In contrast, \textit{batch} arrival occurs when a set of workflows arrives together and is sent to the scheduler to be executed simultaneously \cite{17, 18, 56}. Some articles that claim to adopt an online arrival pattern actually aggregate incoming workflows within a time interval and submit them to the scheduler as a batch \cite{46, 47}. In this review, such approaches are classified under the batch category.

The third category is \textit{ensemble} arrival, which is similar to the batch arrival pattern but with a key distinction: all workflows in an ensemble share the same structure but have different sizes and input data \cite{20, 21, 53, 70, 84}. Users usually specify QoS requirements, such as dedaline and budget, for the entire ensemble, not individual workflows, making scheduling and provisioning more challenging than in batch arrivals. The final category is \textit{periodic} arrival, where multiple workflows are released at regular intervals, with each having a defined period between consecutive instances. Generally, an instance of a workflow must complete before the next one is released, with the release of the next instance acting as a deadline for the previous one. Only two of the reviewed articles address this arrival pattern \cite{1,5}, and both consider using the reserved pricing model for running periodic workflows. Figure \ref{fig:arrival} demonstrates that the online pattern appears most frequently in prior research, with the batch arrival pattern being the second most common.

\subsubsection{Negotiation}
Negotiation is the process where the user and the WaaS broker agree on an SLA, such as pricing or execution time. For instance, in a deadline-constrained case, the user may specify a desired deadline for workflow completion, and the broker and user can then agree on a price. Negotiation can be divided into two types: \textit{static} and \textit{dynamic}. In \textit{static} negotiation, the user specifies their QoS requirements, and the broker checks if it can fulfill them. If it can not, the request is declined, but if it can, the broker sends back a price for the user to accept or reject. However, standards like WS-Agreement Negotiation \cite{battre2010proposal} enable the broker to negotiate QoS and price dynamically with users. For example, in a deadline-constrained scenario, if the broker cannot meet the specified deadline, it might propose a feasible deadline to the user, at a lower price.  

Dynamic negotiation has not been explored in any of the reviewed works; only Taghavi et al. \cite{2} proposed a static negotiation module for a WaaS broker. Their approach considers deadline-constrained workflows: the user submits a workflow with a deadline, and the broker evaluates its feasibility by calculating the critical path. If the deadline is achievable, the broker estimates the execution price based on factors such as the deadline, resource availability, costs, and the broker's profit margin, which is then offered to the user for approval or rejection. While this represents an initial step, negotiation—particularly dynamic negotiation strategies—remains an open and promising research direction (see Section \ref{sec:towardNegotiation}).

\subsubsection{Pricing}
Pricing is the process of determining the cost of the services offered by the WaaS broker. To set the price, the broker must account for the actual cost of executing the workflow, its overhead costs (such as penalties for violating the SLA), its profit margin, and the user's benefits. Typically, the price should be lower than the cost of the user executing the workflow directly by leasing resources from the CSP. This aim can be achieved through efficient reuse of leased resources across multiple workflows, combined with cost-aware provisioning strategies that leverage lower-cost reserved and spot instances.

Pricing can be approached in two ways: \textit{pre-pricing} and \textit{post-pricing}. In pre-pricing, the broker sets the price for executing the workflow before it begins. This involves predicting the execution cost by examining the workflow’s structure, its required QoS (e.g., deadline), and the current availability and costs of resources. Specifically, when using reserved or spot instances, the broker must assess the idle capacity of reserved instances and estimate the future cost of spot instances used for execution. The broker then uses this predicted cost to establish a final price, factoring in hidden costs and its profit. Pre-pricing is also employed during the negotiation phase to finalize the price with the user ahead of execution. In contrast, post-pricing involves determining the final price based on the actual cost incurred during workflow execution. While post-pricing offers greater accuracy and is more beneficial for the broker, users typically prefer knowing the price in advance to decide whether to accept or reject the offer.

Although the majority of the papers reviewed focus on the execution cost of workflows, either by minimizing the cost or staying within the user-defined budget, only one paper addresses the pricing policies of WaaS brokers. Taghavi et al. \cite{2} introduce a post-pricing policy, called \textit{constant profit}, and two pre-pricing policies: \textit{constant discount} and \textit{prediction-based}, for a deadline-constrained WaaS broker that utilizes a combination of on-demand and spot instances for workflow execution. In the constant profit model, after completing the workflow, the broker calculates the total execution cost and adds a fixed profit margin. In the constant discount approach, the broker estimates the workflow’s execution cost using on-demand instances and then applies a discount based on past experience with spot instances before presenting the final price to the user. The prediction-based pricing method involves the broker estimating the cost of execution ahead of time, using the current prices of on-demand instances and the anticipated cost of spot instances during the workflow’s runtime. These pricing strategies highlight a promising area for further development in WaaS brokers.
  
\subsection{Task Scheduling Taxonomy}
Figure \ref{fig:scheduler} illustrates the proposed taxonomy for task scheduler components, comprising seven categories. Each category is detailed in the following sections.

\begin{figure}
    \centering
 
	\begin{forest} 
		for tree={
				font=\scriptsize, 
		    	grow'=east,
		        parent anchor=east,
    		    child anchor=west,
    	        edge path={
                \noexpand\path [draw, \forestoption{edge}] (!u.parent anchor) -- +(10pt,0) |- (.child anchor)\forestoption{edge label};
            }
        }
		[Task Scheduling
			[Task Type [Rigid] [Flexible]]
			[Execution Time Estimation [Fixed] [Probabilistic] [Fuzzy] ]
			[Task Concurrency [Single] [Multiple]]
			[Data Transmission [Direct]	[Indirect]	]
			[Scheduling Criteria [Execution Time] [Monetary Cost] [Resource Utilization] [Energy Consumption] [Fairness] [Security] [Privacy] [Reliability]  ]
			[Scheduling Type [Static [One-by-one] [Combined]] [Dynamic [Immediate] [Periodic]]]
			[Scheduling Algorithm [Heuristic] [Meta-heuristic] [Hyper-heuristic] [Reinforcement Learning] [Meta-learning] [Multiple-Criteria Decision Making] [Mathematical Optimization] [Game Theory]  ]
		]
	\end{forest}

    \caption{Task scheduler taxonomy.}
    \label{fig:scheduler}
\end{figure}

\subsubsection{Task Type}  
The task type defines the degree of flexibility in the task's execution time. Some studies consider tasks to be \textit{rigid}, meaning they require a fixed amount of computational resources, with execution time remaining (nearly) constant on these resources \cite{2, 61}. However, most studies focus on \textit{flexible} tasks, which exhibit varying performance depending on the resources used. To represent flexibility, some studies assume the execution time of each task on each resource is pre-determined through techniques like profiling \cite{7}. Others assume that each task has a known execution time on the fastest resource, with other resources assigned a weight reflecting their relative computational power. Based on this weight, the task's execution time on any resource can be derived \cite{9, 11}. Lastly, some studies estimate the task size in Million Instructions (MI) and the resource's CPU capacity in MIPS (Million Instructions Per Second), calculating the task's execution time by dividing these two values \cite{19, 32}.

\subsubsection{Execution Time Estimation}
Most proposed scheduling algorithms rely on an estimation of each task's execution time. Many studies assume a fixed execution time \cite{28, 37}, which is impractical in dynamic environments like the cloud, where resource performance can be unpredictable. To address this, some researchers adopt probabilistic execution times for workflow tasks, modeling task execution time as a stochastic variable with an independent normal distribution characterized by a specific mean and variance \cite{6, 8, 30}. For instance, Chet et al. \cite{4} model both task execution time and data transfer time among tasks as random variables and use their quantiles for approximation. A few studies manage execution time uncertainty using fuzzy logic. Zhu et al. \cite{16} represent task execution times and data transfer times using triangular fuzzy numbers. Based on these representations, they calculate fuzzy start and completion times for each task. They then introduce a list-based heuristic algorithm, Fuzzy Task Scheduling (FTS), which uses fuzzy priorities to schedule tasks effectively.

\subsubsection{Task Concurrency}
Task concurrency defines the number of tasks that can be executed concurrently on the same resource. Most studies adopt a \textit{single} mode, where only one task is scheduled on each resource at a time \cite{4, 9}. However, some studies explore the \textit{multiple} mode, which permits the concurrent execution of several tasks on the same resource. For instance, in \cite{2, 41}, brokers rent large VMs with a high number of CPU cores and significant memory, bundling tasks together and scheduling them on these VMs as containers. Similarly, Silva et al. \cite{62} utilize multi-threading on a single node to enable the simultaneous execution of multiple tasks.

\subsubsection{Data Transmission}
Data transmission outlines how output data from parent (predecessor) tasks is delivered to child (successor) tasks. In the \textit{direct} (or \textit{peer-to-peer}) approach, data is transferred straight from the resource executing the predecessor task to the resource assigned to the successor task \cite{12, 23}. This method requires synchronous communication, meaning the successor task must already be scheduled, and its resource allocation must be determined beforehand. A notable advantage of this approach is that data transfer time between tasks scheduled on the same resource becomes negligible. However, it may not be ideal for scheduling strategies that delay the execution of successor tasks. In contrast, the \textit{indirect} (or \textit{shared}) method involves the predecessor task storing its output data in shared storage \cite{13, 22}. The successor task can then retrieve this data asynchronously, enabling more flexibility in task scheduling. 

\subsubsection{Scheduling Criteria}
\label{sec:criteria}
The most frequently addressed criteria are the \textit{execution time} of workflows and the rental \textit{cost} of resources, which appear in nearly all studies either as explicit optimization objectives or as constraints (such as deadlines or budgets). In addition, several studies highlight the importance of maximizing rented \textit{resource utilization} \cite{11, 65, 67}, considering it a key measure of the efficiency and effectiveness with which scheduling algorithms leverage available cloud resources.

Another frequently studied objective is the minimization of \textit{energy} consumption. Although energy efficiency is typically considered more of a provider-side concern, a number of studies have explored mechanisms that can be implemented at the broker side to reduce energy usage. One line of research proposes dynamically adjusting the CPU frequencies of leased resources through dynamic voltage and frequency scaling (DVFS) technology \cite{9, 24, 46}, thereby lowering energy consumption during execution. Another approach \cite{32} employs \textit{consolidation}, where tasks are assigned to the minimum number of resources to maximize utilization. In this strategy, tasks are preferentially scheduled on highly utilized resources, enabling underutilized hosts to become idle and transition into sleep (hibernate) mode, ultimately saving energy. Along similar lines, \cite{80} models the energy consumption of each resource as a linear function of its utilization and seeks to maintain utilization at optimized levels to minimize total energy usage. In hybrid cloud environments, where both private and public clouds are used, researchers often emphasize reducing energy consumption in the private cloud while disregarding the rental costs incurred in the public cloud \cite{46, 49}.

\textit{Fairness} has also emerged as an important criterion, though its definition varies across the literature. Some studies \cite{4, 17} treat fairness as a resource-centric property aimed at ensuring balanced workload distribution among resources. In contrast, other works \cite{36, 56, 15, 84} adopt a user-centric perspective, emphasizing equitable resource allocation among workflows. For instance, \cite{36} introduces the concept of a workflow’s \textit{loss rate}, which combines two factors: \textit{slowdown} (the ratio of a workflow’s completion time when scheduled alongside others to its completion time in isolation) and \textit{overspending} (the cost ratio under the same conditions). Fairness is then defined as minimizing disparities in loss rates across workflows. Similarly, Yang et al. \cite{84} propose \textit{slowdown variation} as a fairness metric, measuring the sum of absolute differences between each workflow’s slowdown and the average slowdown across all workflows.

Three studies explicitly addressed \textit{privacy} in hybrid private/public clouds \cite{49, 66, 83}. They define privacy-sensitive tasks as those containing private information, such as financial or medical data, which must be executed on the local private cloud. Ordinary tasks, by contrast, may be executed on either public or private resources. Reference \cite{66} further introduced an intermediate category of tasks that can run on private resources or on a restricted subset of public resources.  

\textit{Security} has also been examined in several studies \cite{34, 73, 82, 83}. These works converge on a similar definition, encompassing three core services: authentication, integrity (via hash algorithms), and confidentiality (via cryptographic algorithms). Together, these services aim to defend against major security threats such as identity spoofing, data alteration, and unauthorized access or snooping. In this framework, each workflow task is assigned a security demand (or level) for each service, ranging from 0 to 1, with higher values reflecting greater sensitivity. Scheduling decisions are then made such that each task is executed on a resource or service whose security level meets or exceeds its requirements.

Another study investigated \textit{trust} \cite{35} as a QoS requirement from the perspectives of both users and CSPs. The authors define trust as the subjective judgment of the trustor, based on personal experience and relevant knowledge, including authenticity, integrity, reliability, and stability. For a user to transact with a CSP, the decision process begins by checking their history of direct interactions with that CSP. If sufficient past transactions exist, the user relies solely on their computed \textit{direct trust}. Otherwise, the user broadcasts a request for recommendations to other trusted entities and calculates an \textit{integrated trust} value by combining limited direct trust with \textit{recommended trust} obtained from others’ opinions. This integrated value forms the basis for the final decision. CSPs, in contrast, adopt a simpler mechanism: they maintain a \textit{blocklist} of known malicious entities and apply the straightforward rule that “everyone is allowed to request service unless they are on the blocklist.”

\textit{Reliability} has been considered in two studies, where it is defined as the probability of completing a task without failure on a given resource. In \cite{9}, reliability is computed from the resource’s fault rate and operating frequency, and the authors employ \textit{task duplication} to improve the reliability of tasks deemed less reliable. In contrast, \cite{43} explores the joint provisioning of spot and on-demand instances. In this model, the reliability of executing a task on a spot instance is estimated through a cumulative distribution function, which approaches unity as the task’s finish time moves further from the instance’s launch time—reflecting the observation that freshly launched spot instances have lower failure probabilities than older ones. The authors then aim to strike a balance among time, cost, and reliability when scheduling workflow tasks.

\subsubsection{Scheduling Type}
Workflow scheduling can be classified into two primary types: \textit{static} and \textit{dynamic} \cite{rodriguez2017taxonomy}. Static scheduling involves assigning tasks to resources prior to the start of workflow execution. This approach typically produces high-quality solutions since it considers crucial structural aspects of the workflow, such as the critical path. However, static scheduling is limited by its reliance on precise runtime estimations and its inability to adapt to unforeseen changes in resource availability, making it less suitable for the uncertain nature of cloud environments. On the other hand, dynamic scheduling makes decisions during runtime, assigning tasks as they become \textit{ready}—that is, once all their predecessor tasks have been completed. This enables dynamic methods to respond effectively to changes and uncertainties in the cloud environment. Nevertheless, these algorithms often make greedy, task-specific decisions, lacking a global perspective on the entire workflow. Consequently, the solutions generated by dynamic approaches are generally less optimal compared to those of static methods. To address the shortcomings of both approaches, hybrid strategies have been proposed. These methods begin by creating a static execution plan for the workflow and then apply dynamic adjustments during runtime to account for environmental uncertainties. This combination leverages the strengths of both static and dynamic scheduling to produce more robust solutions.

In static multiple workflow scheduling, two primary approaches have been used: \textit{one-by-one} and \textit{combined}. In the one-by-one method, workflows are scheduled sequentially, based on priority, while considering resource availability \cite{15, 45}. For example, Rezaeian et al. \cite{15} apply the one-by-one approach using four different workflow selection methods: Ordered, which schedules workflows based on a predefined priority; round-robin, which iterates over workflows, scheduling one task from each workflow at a time; weighted round-robin, where workflows are selected according to a weighted formula before scheduling a task; and unfairness heuristic, which selects workflows based on their level of unfairness. In contrast, the \textit{combined} approach merges all workflows into a single large workflow before scheduling it on available resources. References \cite{34, 40} use a similar approach, where they break down each workflow into levels based on task depth. Then, they create partitions where each partition contains tasks at the same level across different workflows. Other studies merge workflows by adding a dummy entry task to connect the input workflows' actual entry tasks, and a dummy exit task to connect the workflows' exit tasks \cite{1, 44, 69}.

Two strategies are commonly used for dynamic multiple workflow scheduling: \textit{immediate} and \textit{periodic}. In the immediate strategy, tasks are scheduled as soon as they become \textit{ready}. However, this approach faces two challenges: first, the runtime overhead caused by executing the scheduling algorithm for each task as it becomes ready, and second, the greedy decision-making process for each task without considering the future tasks and their potential impact on the current scheduling. To address these issues, the periodic strategy was introduced. This method divides time into intervals, collecting all ready tasks within each interval and scheduling them all at once at the end of the interval. While this approach resolves the problems associated with immediate scheduling, it also introduces a delay, causing ready tasks to wait until the end of the interval, potentially missing deadlines. Most previous work in dynamic scheduling adopts the immediate strategy \cite{8, 6, 13}, although some utilize the periodic strategy \cite{2, 74}. Chen et al. \cite{42} introduced a hybrid strategy, where they use periodic scheduling but select tasks at the start of each interval that are predicted to have their earliest start time within the current interval. These tasks are scheduled on available resources ahead of time.

Four studies have employed a combination of both static and dynamic scheduling types \cite{1, 55, 61, 86}. Taheri et al. \cite{1} proposed a framework consisting of two phases: static planning and dynamic scheduling. In the static planning phase, a heuristic algorithm is used to schedule a combined workflow. During the dynamic scheduling phase, the system seeks to provision the necessary resources and execute tasks based on the static plan. However, adjustments may be made to the initial schedule to account for uncertainties in the cloud environment. Similarly, Zhou et al. \cite{55} introduced a comparable framework with an offline optimization stage using breadth-first search or A* algorithms to create a static schedule, followed by an online adaptation stage that refines the primary schedule using runtime information about cloud resources and workflow executions. Finally, Toporkov et al. \cite{86} perform a static analysis of each incoming workflow using the Critical Jobs’ Method (CJM) to determine feasible execution windows for all tasks. These windows are then used to group tasks into independent execution batches that respect data dependencies. At runtime, as each batch approaches its planned start time, a dynamic assignment process selects and provisions suitable VMs by solving a matching problem.

\begin{figure}[t] % "[t!]" placement specifier just for this example
	\begin{subfigure}{0.49\linewidth}
	\includegraphics[width=\linewidth]{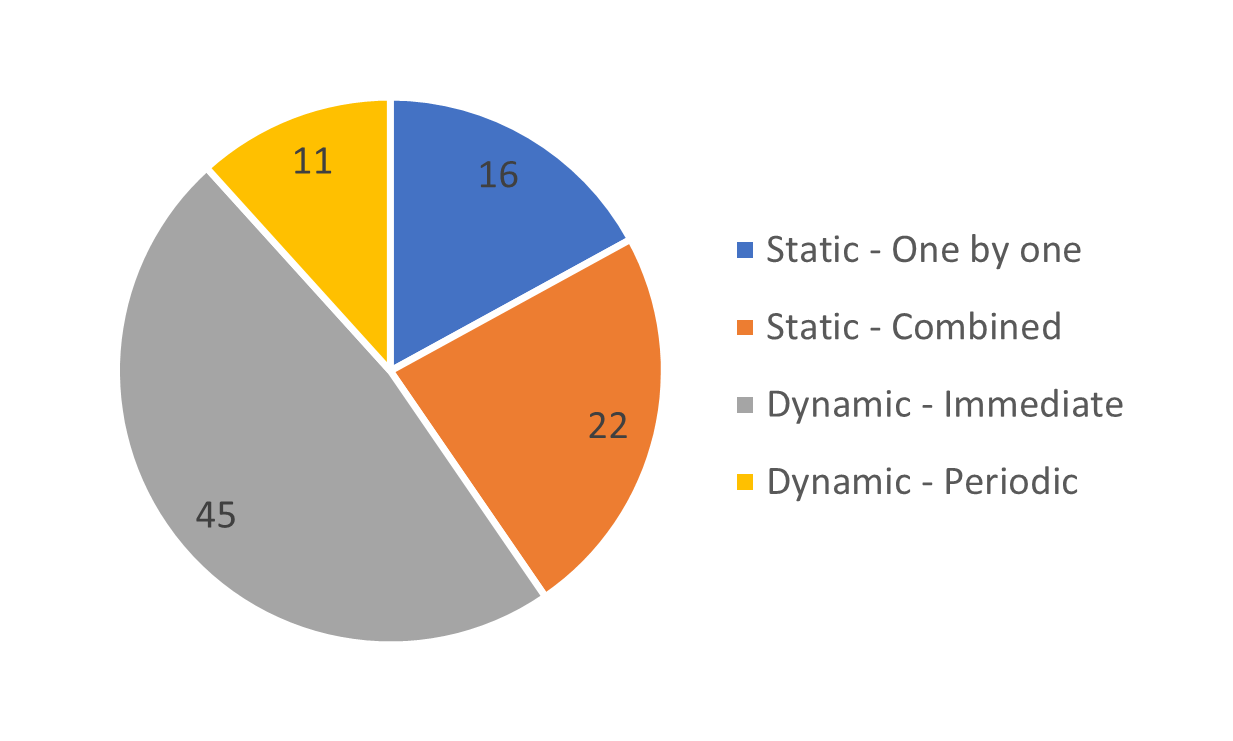}
	\caption{Scheduling types}\label{fig:types}
	\end{subfigure}
	\begin{subfigure}{0.49\linewidth}
	\includegraphics[width=\linewidth]{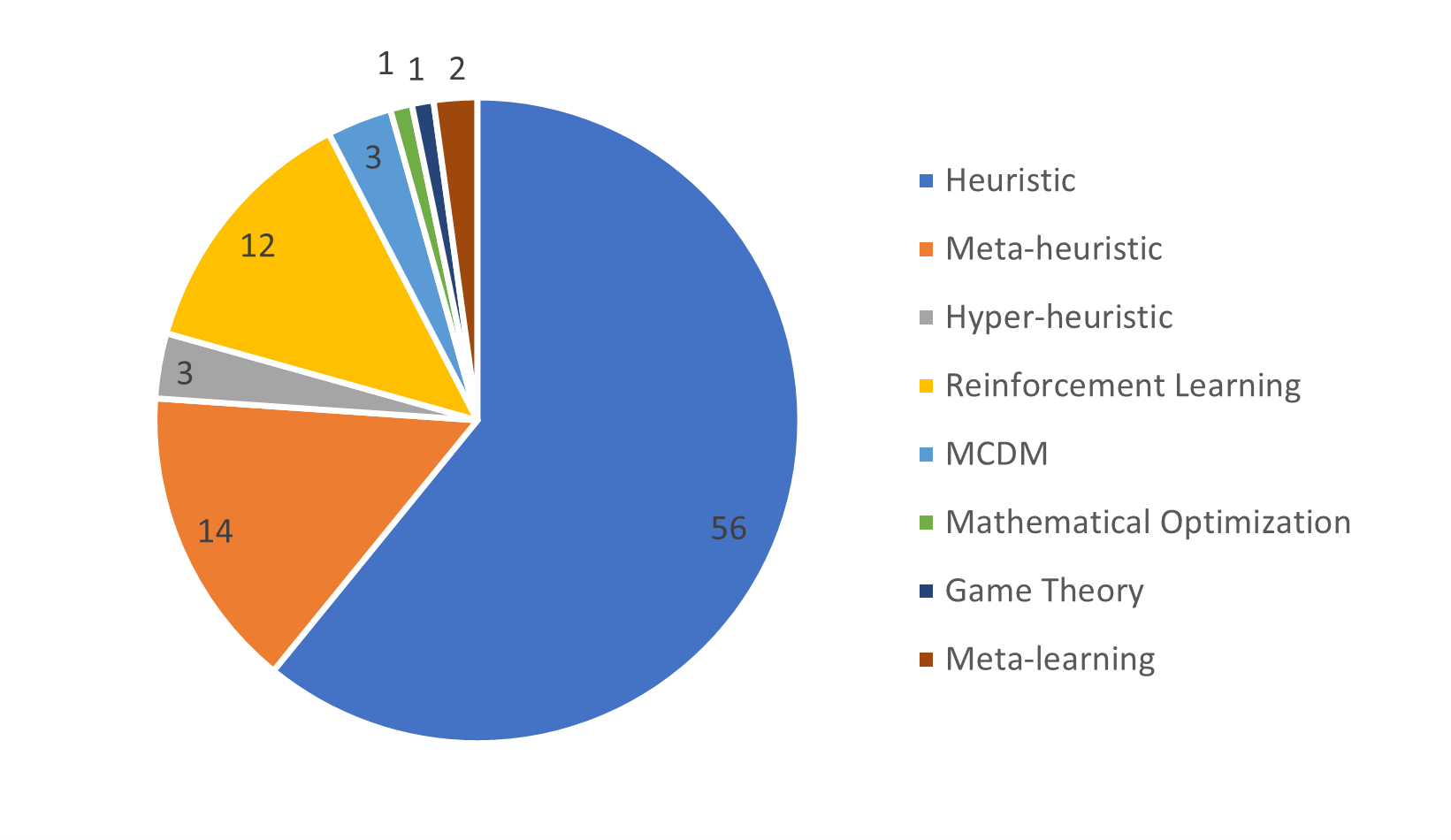}
	\caption{Scheduling algorithms}\label{fig:algorithms}
	\end{subfigure}\hspace*{\fill}
	\caption{Number of articles for different scheduling types and algorithms} 
	\label{fig:schMethods}
\end{figure}

Figure \ref{fig:types} illustrates the distribution of articles across each category. It is evident that immediate dynamic scheduling is the most commonly used approach.

\subsubsection{Scheduling Algorithm}
\label{sec:techniques}
The MWSP in a cloud environment is a well-known NP-hard problem \cite{ulman1975np}. Various algorithms have been employed by researchers to address this issue, with \textit{heuristic} methods being the most commonly applied. The term \textit{heuristic} refers to a method based on experience or judgment, designed to provide a solution that is close to optimal for a given problem \cite{silver2004overview}. The majority of these studies have utilized the \textit{list heuristic} to solve the MWSP \cite{4, 14, 52, 23, 68, 38, 75}. This heuristic involves two stages: task selection and resource mapping. In the task selection phase, each workflow task is assigned a rank according to a specific formula, after which the tasks are sorted and selected one by one based on their ranks. In the resource mapping phase, each selected task is assigned to a resource that optimizes the scheduling objective. Another commonly used heuristic in this field is the \textit{clustering heuristic} \cite{36,40}, where tasks of the workflow are grouped into clusters, and each cluster is scheduled on a single resource to minimize data transmission time between tasks (in direct transmission mode). Other heuristic algorithms, such as A*, have also been explored by researchers \cite{54,55}.

\textit{Meta-heuristic} methods are another popular class of algorithms for task scheduling. Meta-heuristics are higher-level heuristics that are problem-independent and provide generic frameworks for finding near-optimal solutions to optimization problems \cite{sangaiah2018metaheuristic}. Li et al. \cite{3} proposed the Multi-swarm Co-evolution-based Hybrid Intelligent Optimization (MCHO) algorithm, which aims to minimize makespan and cost while meeting the deadline of each workflow. They utilized the Particle Swarm Optimization (PSO) algorithm, incorporating three different swarms: the first and second swarms focus on optimizing makespan and cost, respectively, while the third swarm is introduced to balance the two objectives. Additionally, they employed Genetic Algorithms (GA) as an elite enhancement strategy for each swarm, aiming to exploit elite individuals and increase the likelihood of finding more non-dominated solutions. Non-dominated Sorting Genetic Algorithm III (NSGA-III) is also applied to address the MWSP as a prominent multi-objective meta-heuristic method \cite{36,56}. Li et al. \cite{36} first introduced a clustering heuristic called Depth-First-Search considering Critical Subsequent Task (DFS-CST) to partition each workflow into clusters and then applied NSGA-III to assign resources to each cluster, aiming to minimize three objectives simultaneously: execution time, execution cost, and unfairness.

Qin and Shao \cite{44} proposed a bi-population co-evolutionary algorithm for multi-objective, deadline-constrained multi-workflow scheduling. Their method involves two sub-populations: a deadline-aware sub-population that focuses on finding feasible schedules that meet deadlines, and a multi-objective sub-population that aims to minimize both makespan and cost. After each iteration, a bi-directional exchange of schedules occurs between the two sub-populations, promoting the diversity of each population. Other meta-heuristic algorithms, such as the Simulated Annealing (SA) \cite{87}, Owl Search Algorithm (OSA) \cite{46}, Poor and Rich Optimization (PRO) algorithm \cite{47}, the Snake Optimizer (SO) \cite{48}, the Gradient-Based Optimizer (GBO) \cite{50}, and the Linear Congruential Lyrebird Optimization Algorithm (LC-LOA) \cite{77} have also been applied to address the MWSP. 

\textit{Hyper-heuristics} have also been applied to address MWSP. Unlike meta-heuristics, which directly search the solution space, hyper-heuristics operate at a higher level by selecting, combining, generating, and sequencing problem-specific heuristics from a predefined set of heuristics and/or meta-heuristics \cite{dokeroglu2024hyper}. Their search is conducted in the space of heuristics rather than the space of problem solutions. In particular, three studies have employed Genetic Programming Hyper-Heuristics (GPHH) to evolve heuristic rules for multi-workflow scheduling \cite{39, 57, 72}. In this approach, each heuristic is represented as a syntax tree (also known as a GP tree), where the terminal nodes correspond to problem parameters—such as task execution time, or the execution speed and cost of a VM—while the internal nodes represent function operators, such as addition or maximum. Each tree represents a candidate heuristic evolved by GPHH, and the most effective heuristic(s) are ultimately selected for scheduling.

Recently, \textit{Reinforcement Learning (RL)} has been applied to solve the MWSP. RL is a machine learning approach where an agent learns through interaction with an environment, making decisions based on feedback from its actions and experiences. The agent selects actions based on its policy and the environment's state, receives rewards or penalties as feedback, and adjusts its policy accordingly. Li et al. \cite{33} use a Double Deep Q-Network (DDQN) to minimize both makespan and cost in a multi-workflow scheduling context. They employ a two-level scheduling strategy: at the first level, a ready task is selected for scheduling, and at the second level, a suitable \textit{VM} is chosen for that task. Each level features two sub-agents: a time sub-agent and a cost sub-agent, aimed at optimizing time and cost, respectively. After training, the model achieves an optimal balance between makespan and cost after 400 episodes. Pan and Wei \cite{59} proposed a similar dynamic scheduling algorithm using a Deep Q-Network (DQN), but focusing solely on the VM selection phase. In another related study, Huang et al. \cite{71} utilize a Deep Neural Network (DNN) for VM selection of the ready tasks, combined with an evolutionary algorithm to optimize the DNN's parameters. 

Chen et al. \cite{78} proposed Arbitrary-Freedom Adaptive Scheduling (AFAS), which also employs Deep Reinforcement Learning (DRL) to prioritize \textit{ready tasks} and generate a scheduling sequence for mapping them to available resources. In their approach, workflow tasks are categorized into two types: \textit{routine tasks}, which appear frequently with predictable patterns and resource requirements, and \textit{nonroutine tasks}, which occur irregularly and exhibit variable resource demands. The proposed adaptive scheduling algorithm combines a heuristic-based scheduling method for handling nonroutine tasks with a DRL-based intelligent scheduling strategy tailored for routine tasks. In another study, He et al. \cite{83} first employed a Q-learning approach to assign priorities to the ready tasks, and then used a DRL method to map the prioritized tasks to available VMs.

Lin et al. \cite{74} also employed RL for workflow task scheduling across a combination of on-demand and spot instances, though in a different way. For task scheduling, they used a DNN that takes an 11-dimensional vector of the ready task's features as input. Instead of selecting a specific VM for task execution, the output is a meta-policy. The meta-policies are as follows: \textit{Util-priority}: schedule the task on the available spot instance with the least available resources; \textit{Avl-priority}: schedule the task on newer spot instances; \textit{To-spot}: launch a new spot instance and schedule the task on it; \textit{To-on-demand}: schedule the task on an idle available on-demand instance, or if none are available, launch a new on-demand instance; \textit{Wait}: defer the task's execution until the next scheduling period. While all the RL-based approaches focus on dynamic scheduling, Zhang et al. \cite{60} proposed a static scheduling algorithm for handling multiple workflows. Their approach first utilizes GA to determine the optimal ordering of the workflow's tasks, and then employs a DQN agent to assign the whole workflow to a VM instance. 

To prepare richer feature vectors for workflow tasks and resources, several studies have employed \textit{Graph Neural Networks (GNNs)} \cite{zhou2020graph}. GNNs are a class of deep learning models designed to operate on graph-structured data (consisting of nodes and edges). They work by aggregating information from a node’s neighbors and updating the node’s representation based on its own features and the features of its neighboring nodes. This mechanism allows GNNs to capture both the attributes of individual nodes and the overall graph topology. The output of a GNN is typically an \textit{embedding}—a numerical vector representation of a node (or edge or entire graph)—that encapsulates its properties and structural context. These embeddings can then be used as input features to a reinforcement learning-based method.

For example, Wang et al. \cite{79} employed a Graph Attention Network (GAT)—a specific type of GNN—to encode the features of each ready task along with its neighboring tasks. They applied a similar strategy to generate embeddings for each VM in the network. The concatenation of these task and VM embeddings formed the input to a DRL algorithm, which was then used to schedule ready tasks on the available resources. Along the same line, Huang et al. \cite{81} also integrated GAT with DRL, but their objective was specifically to minimize execution costs while meeting deadline. Finally, Chandrasiri and Meedeniya \cite{80} proposed a method that constructs a heterogeneous graph whose nodes include both workflow tasks and VMs. A GNN is then used to compute embeddings for this combined graph. These embeddings serve as inputs to a DRL-based scheduler that makes task-to-VM assignment decisions with the dual objective of minimizing both makespan and energy consumption.

\textit{Meta-learning}, often described as \textit{learning to learn,} is a paradigm where algorithms are trained not just to solve individual learning problems but also to acquire the ability to adapt quickly to new ones \cite{vettoruzzo2024advances}. Instead of starting from scratch for each new scheduling scenario, a meta-learner captures transferable knowledge across related scheduling problems and uses it to accelerate adaptation to unseen workflows. In the context of multiple workflow scheduling, this means the scheduler can exploit similarities between workflows—such as recurring DAG structures, resource contention patterns, or typical deadline/cost trade-offs—to refine strategies more efficiently. 

One branch of meta-learning is Evolutionary Multi-Task Optimization (EMTO), which evolves scheduling heuristics or policies simultaneously across several scheduling problems. In this way, insights gained in optimizing one workflow (e.g., dealing with long critical paths or VM bottlenecks) can be transferred across workflows, reducing search costs and improving overall efficiency. This knowledge transfer between scheduling problems is particularly valuable in WaaS environments, where workflows often arrive with diverse characteristics but share common structural or performance patterns. Zhou et al. \cite{18} employed EMTO to address deadline-constrained multi-workflow scheduling, while Zhang et al. \cite{76} applied a similar approach to develop a scheduling algorithm that simultaneously minimizes makespan, cost, and energy consumption. Another widely studied meta-learning approach is meta-reinforcement learning (meta-RL) \cite{beck2025tutorial}, which has been applied to the scheduling of multiple workflows in hybrid edge–cloud environments \cite{chen2025dynamic, zhen2025multi}. However, its application in cloud-only settings has not yet been investigated, making it a promising direction for future research (see Section \ref{sec:AI-driven}).

\textit{Multi-criteria decision-making (MCDM)} is a field that focuses on choosing the best alternative from several options by evaluating multiple criteria or attributes \cite{taherdoost2023multi}. Among the popular MCDM methods, only the Technique for Order of Preference by Similarity to Ideal Solution (TOPSIS) has been applied to the MWSP \cite{10, 26, 29}. Chakravarthi et al. \cite{26, 29} use TOPSIS to determine the best VM for a ready task in a dynamic scheduling algorithm. They input a $m \times r$ decision matrix into TOPSIS, where $m$ represents the number of available VMs, and $r$ is the number of criteria (3 in this case): execution time, transfer time, and execution cost. 

\textit{Game theory} is another approach used to model scenarios in which multiple independent players interact and influence each other's outcomes \cite{nisan2007algorithmic}. Among the reviewed studies, only one uses this method. Wu and Yuandou \cite{17} apply multi-stage dynamic game theory to minimize makespan while enhancing fairness and utilization. Each optimization objective—makespan, fairness, and utilization—is considered as a competing player. The Nash equilibrium is determined as the optimal solution to the problem.

Finally, \textit{mathematical optimization} (or mathematical programming) utilizes techniques like Linear/Nonlinear Programming (LP/NLP) to address optimization problems. Karmakar et al. \cite{58} apply NLP to minimize execution costs for a set of deadline-constrained workflows. They first identify the critical paths within the workflow and then aim to minimize the total required MIPS for tasks on these critical paths, ensuring the workflow’s deadline is met. The problem is formulated as an NLP, and an approximate solution is found using the Lagrange Multiplier Method. Figure \ref{fig:algorithms} shows the distribution of articles for each algorithm. It is clear that heuristic algorithms are the most frequently used, followed by meta-heuristic algorithms.

\subsection{Resource Provisioner Taxonomy}
\begin{figure}
    \centering
 
	\begin{forest} 
		for tree={
		    	grow'=east,
	        parent anchor=east,
    		    child anchor=west,
            edge path={
                \noexpand\path [draw, \forestoption{edge}] (!u.parent anchor) -- +(10pt,0) |- (.child anchor)\forestoption{edge label};
            }
        }
		[Resource Provisioner, l sep=5mm, 
			[Resource Type, l sep=5mm,[Virtual Machine] [Container] [Function] [Quantum Machines]]
			[Resource Diversity, l sep=5mm, [Heterogeneous] [Homogeneous]]
			[Resource Pricing, l sep=5mm, [On-demand] [Spot] [Reserved] ]
			[Provision Timing, l sep=5mm, [As-needed] [Periodic]]
			[Provisioning Delay, l sep=5mm, [Delay-aware] [Instantaneous] ]
			[Deployment Model, l sep=5mm, [Single-Cloud] [Multi-Cloud] [Hybrid Cloud]]
		]
	\end{forest}

    \caption{Resource provisioner taxonomy.}
    \label{fig:provisioner}
\end{figure}

Figure \ref{fig:provisioner} depicts the proposed taxonomy for the resource provisioner component, which is organized into six categories. A detailed explanation of each category follows.

\subsubsection{Resource Type}
\begin{figure}[t] % "[t!]" placement specifier just for this example
	\begin{subfigure}{0.47\linewidth}
	\includegraphics[width=\linewidth]{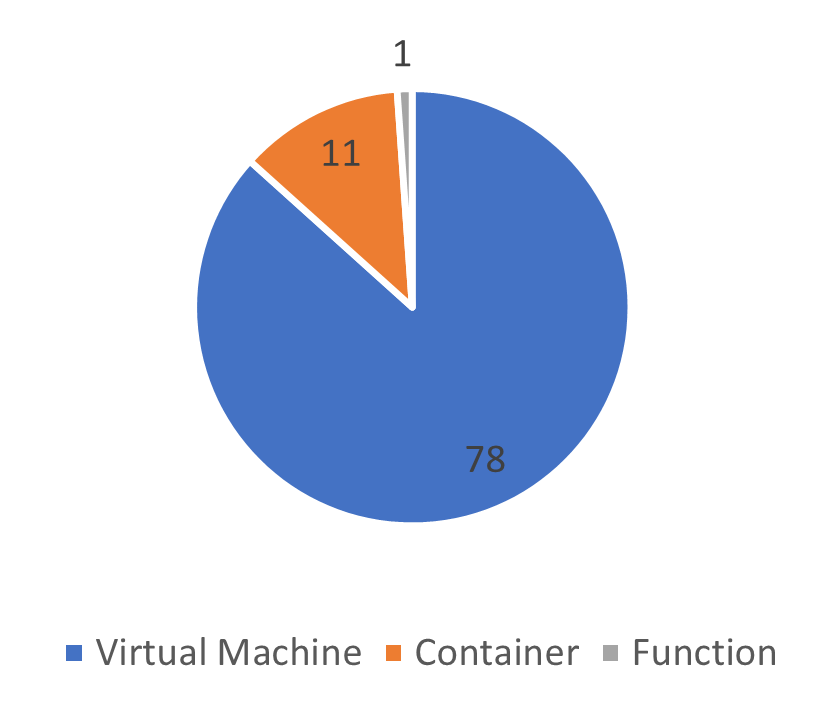}
	\caption{Resource Type}\label{fig:resources}
	\end{subfigure}\hspace*{\fill}
	\begin{subfigure}{0.5\linewidth}
	\includegraphics[width=\linewidth]{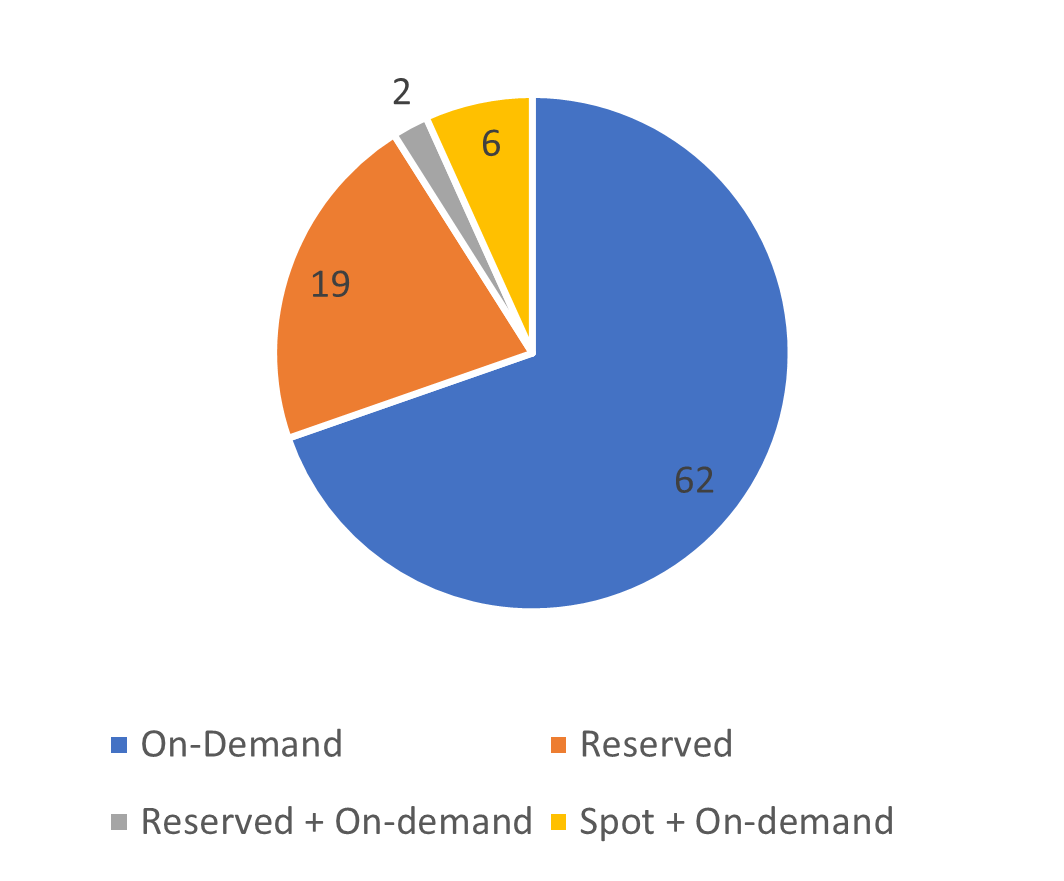}
	\caption{Resource Pricing}\label{fig:pricings}
	\end{subfigure}
	\caption{Number of articles for different resource types and pricing} \label{fig:resprice}
\end{figure}

As illustrated in Figure \ref{fig:resources}, the majority of reviewed articles schedule workflow tasks directly on VMs \cite{51}, which represent traditional IaaS resources. However, VMs are heavyweight, and their high provisioning delay can significantly reduce the efficiency of scheduling algorithms—especially in deadline-constrained scenarios, where such delays may result in missed deadlines. To mitigate provisioning overhead, brokers often keep idle VMs available for future use, which leads to reduced resource utilization—particularly in the case of short- and medium-duration tasks. In response to these limitations, containers have recently gained popularity as a lightweight virtualization technology. Containers offer near-instant provisioning and allow users to encapsulate tasks as isolated, self-contained applications. As a result, some studies have adopted containers to schedule multiple tasks on the same VM, improving utilization while avoiding provisioning delays. For example, in \cite{2, 3}, the broker rents large VMs with many CPU cores and substantial memory, then uses containers to bundle and execute several tasks concurrently on a single VM.

Although serverless computing and FaaS have emerged as promising models for executing scientific workflows in the cloud \cite{malawski2020serverless}, only Sun et al. \cite{57} have investigated the scheduling of multiple workflows using a hybrid approach that combines serverless functions and VMs. Their study was conducted in a multi-cloud environment spanning AWS, GCP, and Azure. They applied GPHH hyper-heuristic (see Section \ref{sec:techniques}) to automatically generate three heuristic rules for: (1) task selection, (2) CSP selection, and (3) resource selection. The third rule accounts for both VMs and serverless functions when selecting resources to execute tasks, aiming to minimize the total execution cost while satisfying deadline constraints. The combination of serverful resources and serverless functions shows significant potential, which is discussed as a promising future direction in Section \ref{sec:hybridServerless}.

To date, no prior work has addressed the scheduling of multiple workflows in hybrid quantum/classical environments, which is unsurprising given the early developmental stage of quantum computing and quantum algorithms. The sole exception is Qonductor \cite{giortamis2024orchestrating}, a WMS designed specifically for multiple hybrid workflows. However, as of mid-2025, it remains an arXiv preprint and has not appeared in a peer-reviewed journal. We discuss this emerging and promising area of hybrid scientific workflows, with particular attention to Qonductor, in the future directions outlined in Section \ref{sec:hybridQuantum}.

\subsubsection{Resource Diversity}
The majority of reviewed studies considered CSPs offering \textit{heterogeneous} resources with varying configurations, such as CPU cores, memory, and storage, aiming to select the optimal resource configuration for each task based on its requirements and scheduling objectives. Conversely, some researchers have opted for a \textit{homogeneous} environment, using a single resource type for all tasks to simplify scheduling and reduce the search space \cite{11, 16}. Malawski et al. \cite{21} justify this approach by noting that, for most applications, typically one or two VM types provide the best price/performance ratio. Similarly, Taheri et al. \cite{1} utilized a single type of large VM with sufficient resources and employed containers to schedule multiple tasks on the same VM.

\subsubsection{Resource Pricing}
As shown in Figure \ref{fig:pricings}, most of the reviewed studies employed the widely used \textit{on-demand} pricing model to execute users’ workflows, taking advantage of the elasticity offered by cloud resources \cite{25, 63}. In contrast, some studies, either implicitly or explicitly, relied on a fixed pool of \textit{reserved} resources, which does not fully reflect the dynamic nature of cloud computing \cite{59, 60}. Nevertheless, Chen et al. \cite{5} utilized reserved resources for executing periodic workflows and proposed a heuristic algorithm to determine the optimal number of reserved instances, leveraging the flexible nature of reserved resources.

More interestingly, several studies explored how the WaaS broker can reduce execution costs by combining \textit{spot} and \textit{on-demand} instances \cite{2, 20, 43, 53, 54, 74}, or \textit{reserved} and \textit{on-demand} instances \cite{1, 16}. Zolfaghari et al. \cite{20} developed a multi-class workflow scheduling algorithm, where tasks are assigned to different resource classes based on their priority. Higher-priority tasks are sent to higher-class resources, which are reliable but expensive on-demand instances, while lower-priority tasks are allocated to cheaper spot instances with lower bid prices. If a task fails to execute within a specified time, it is moved to a higher-class resource until completion. Similarly, Zhou et al. \cite{54} combined spot and on-demand resources for workflow execution. They initially used the A* algorithm to identify the optimal on-demand instance type for each task to minimize costs while meeting the workflow’s deadline. Then, a refinement algorithm replaces some on-demand instances with spot instances to further reduce costs. Taheri et al. \cite{1} focused on scheduling periodic workflows using a mix of reserved and on-demand instances. Their scheduling algorithm prioritizes reserved instances for each task until its sub-deadline is reached. If resources are insufficient, the algorithm switches to on-demand instances to avoid missing the sub-deadline. The combination of different pricing models to reduce execution costs represents a promising strategy, which is further discussed in the future directions in Section \ref{sec:economist}.

\subsubsection{Provision Timing}
Provision timing is categorized into \textit{as-needed} and \textit{periodic} approaches. In \textit{as-needed} provisioning, the broker acquires new resources when a task (or a group of tasks) requires them—typically because they cannot be scheduled on the currently available resources without violating an optimization criterion or constraint (e.g., a sub-deadline). In this approach, the resource provisioner is usually integrated within the task scheduler rather than being a separate module. Whenever there is a resource shortage preventing the scheduling of a task, the algorithm triggers commands to lease new resources. The key advantage of this method is its prompt response to environmental events. For instance, in reference \cite{1}, the scheduling algorithm calculates a parameter called \textit{laxity} for each task, indicating the amount of time it can be delayed before missing its sub-deadline. If a task is ready to execute but no idle resources are available, the scheduler postpones it until the necessary resources become free, provided its laxity is still positive. If the required resources are not released and the laxity reaches zero, the scheduler proceeds to request new resources.

In contrast, \textit{periodic} provisioning, commonly used in dynamic scheduling algorithms, requires the task scheduler to allocate workflow tasks to currently available resources, without actively managing resource provisioning. The resource provisioner operates as an independent module, periodically assessing the current resources based on parameters such as CPU load, task queue size, and other metrics. If necessary, the provisioner decides whether to lease additional resources or release some existing ones. One drawback of this approach is that it reacts to changes in the environment and workflow requirements with a delay, typically only at the end of each time interval. This lag can risk violating certain constraints, such as deadlines, forcing the broker to adopt more conservative decisions in some cases. On the other hand, this delay can also be advantageous. In the as-needed approach, resources are granted immediately whenever a task requests them, often without considering the state of other running or pending tasks. This can lead to premature resource allocation, whereas shortly thereafter a suitable resource might become available, or it might be preferable to wait and rent more powerful resources as new tasks arrive. By deferring resource provisioning until the end of each interval, the broker can make more informed decisions, taking into account the overall status of both resources and queued tasks.

Taghavi et al. \cite{2} combine dynamic periodic scheduling with periodic provisioning, where the provisioning intervals are significantly longer than the scheduling intervals (ranging from 5 to 10 times longer). During each provisioning interval, the provisioner checks whether the number of idle cores on the VMs is sufficient to accommodate the required cores for the ready tasks. If not, new VMs are leased from the CSP. In contrast, Malawski et al. \cite{21} implement dynamic immediate scheduling with periodic provisioning. Their approach computes resource utilization at the end of each interval and adjusts the number of leased VMs if the utilization exceeds or falls below certain thresholds. Tarafdar et al. \cite{32} employ a different approach to periodic provisioning, where, at the end of each time interval, they attempt to predict the number of workflows arriving in the next interval using historical data from previous intervals and scale the resources up or down accordingly.

\subsubsection{Provisioning Delay}
When a broker requests the provisioning of a VM from a CSP, it does not become available immediately. Instead, it takes a significant amount of time to be deployed on a physical host and for its operating system to boot. This period is referred to as \textit{provisioning delay}. Even though containers, a lightweight virtualization technology, have a much shorter provisioning delay than VMs, they still experience some delay. This provisioning delay can be substantial, even in the case of containers, and thus, to make accurate provisioning and scheduling decisions, this parameter should be considered. In particular, provisioning delay plays a crucial role in provisioning algorithms, as in some situations it may be better not to release an idle resource so that it can be used for future tasks, avoiding the delay associated with provisioning new resources. To minimize VM provisioning delays, some researchers opt to use larger VMs and run multiple tasks on these VMs using container technology \cite{2}. However, only about 30\% of the reviewed articles considered provisioning delay \cite{31, 46}, which we refer to as \textit{delay-aware}, while the remainder of the articles did not mention it at all \cite{38, 50}, and are categorized as \textit{instantaneous}.

\subsubsection{Deployment Model}
\label{sec:deployment}
This taxonomy examines how the broker acquires its required resources, either from a single CSP or multiple CSPs. The majority of articles (76 papers) utilize a \textit{single-cloud} provider to lease their resources. However, relying on a single CSP has its drawbacks, including vendor lock-in, limited resource diversity, restricted pricing models, provider unavailability, and geographical location constraints \cite{khorasani2024cloud}. As a result, some studies (7 papers) adopt the \textit{multi-cloud} model, where the broker leases resources from multiple, independent CSPs \cite{9, 30, 51, 57, 55, 35, 17}. In contrast, a few articles (4 papers) implement a \textit{hybrid-cloud} approach, where the broker uses resources from both a local private cloud and a public cloud, turning to the latter when local resources are insufficient \cite{46, 49, 66, 83}. 

\subsection{Monitoring and Fault Tolerance Taxonomy}
\begin{figure}
    \centering
 
	\begin{forest} 
		for tree={
	    		edge path={\noexpand\path[\forestoption{edge}] (\forestOve{\forestove{@parent}}{name}.parent anchor) -- +(0,-12pt)-| (\forestove{name}.child anchor)\forestoption{edge label};}
		}
		[Monitoring and Fault Tolerance
			[Re-execution] [Checkpointing] [Replication] 
		]
	\end{forest}

    \caption{Monitoring and fault tolerance taxonomy.}
    \label{fig:fault}
\end{figure}

Resource failures are an inevitable aspect of computer datacenters, making it crucial for WaaS brokers to monitor resource performance and respond appropriately when failures occur. However, this requirement has not been adequately addressed in many reviewed studies. This is especially important in the spot pricing model, where the unreliability of resources and the potential for CSPs to reclaim them during task execution are concerns. Despite these challenges, only 13 articles, including all 6 that utilize spot instances, address fault tolerance.

Figure \ref{fig:fault} outlines a taxonomy for monitoring and fault tolerance. The simplest method, \textit{re-execution}, involves transferring tasks from a failed resource to a new one, but this results in the loss of all progress made up to that point \cite{21, 37, 43, 54}. This can be costly for the WaaS broker, particularly for scientific workflows with long-running tasks and strict deadlines. To minimize this cost, some papers use \textit{checkpointing} \cite{2, 18, 20, 30}, where snapshots of tasks or VMs are periodically saved. If a failure occurs, the task is reassigned to a new instance, and the most recent snapshot is restored. Additionally, most CSPs provide advance notice before stopping or terminating spot instances, allowing sufficient time to take a snapshot. However, the process of taking and restoring snapshots can be time-consuming.

The final method is \textit{replication}, which involves running duplicate instances of each task across multiple VMs to enhance the reliability of task execution \cite{9}. In the event of a failure, the task can continue running without disruption, as its duplicate instances remain operational on other resources. While this approach ensures task continuity without losing time, it adds additional costs for the broker due to the need for extra resources to run the duplicated tasks. Chen et al. \cite{53} combine replication and re-execution for workflows running on a mix of on-demand and spot instances. They schedule tasks on spot instances only if there is sufficient time for re-execution in case the resource is reclaimed by the provider. Otherwise, the task is scheduled on an on-demand instance. When tasks are assigned to spot instances, they attempt to identify idle time on on-demand instances to replicate the tasks where possible. If a spot instance is reclaimed, tasks running on it are checked for duplicates. Tasks with duplicates are ignored, while those without duplicates are re-executed on another instance.

\subsection{Insights from Previous Work: Trends and Trade-offs}
\begin{figure}[t!] % "[t!]" placement specifier just for this example
	\begin{subfigure}{0.5\linewidth}
	\includegraphics[width=\linewidth]{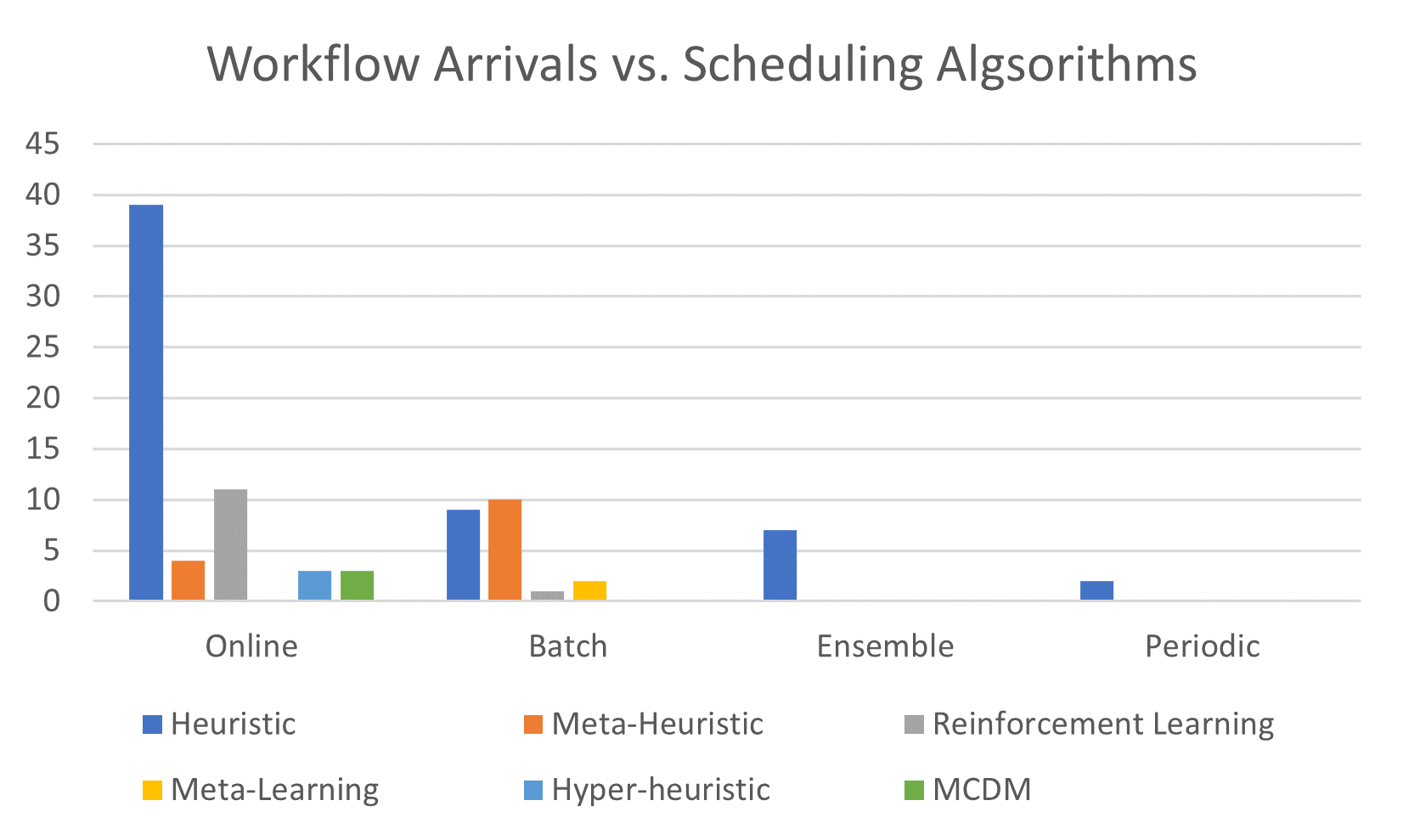}
	%\caption{Workflow Arrivals vs. Scheduling Algorithms}\label{fig:arrAlg}
	\end{subfigure}\hspace*{\fill}
	\begin{subfigure}{0.5\linewidth}
	\includegraphics[width=\linewidth]{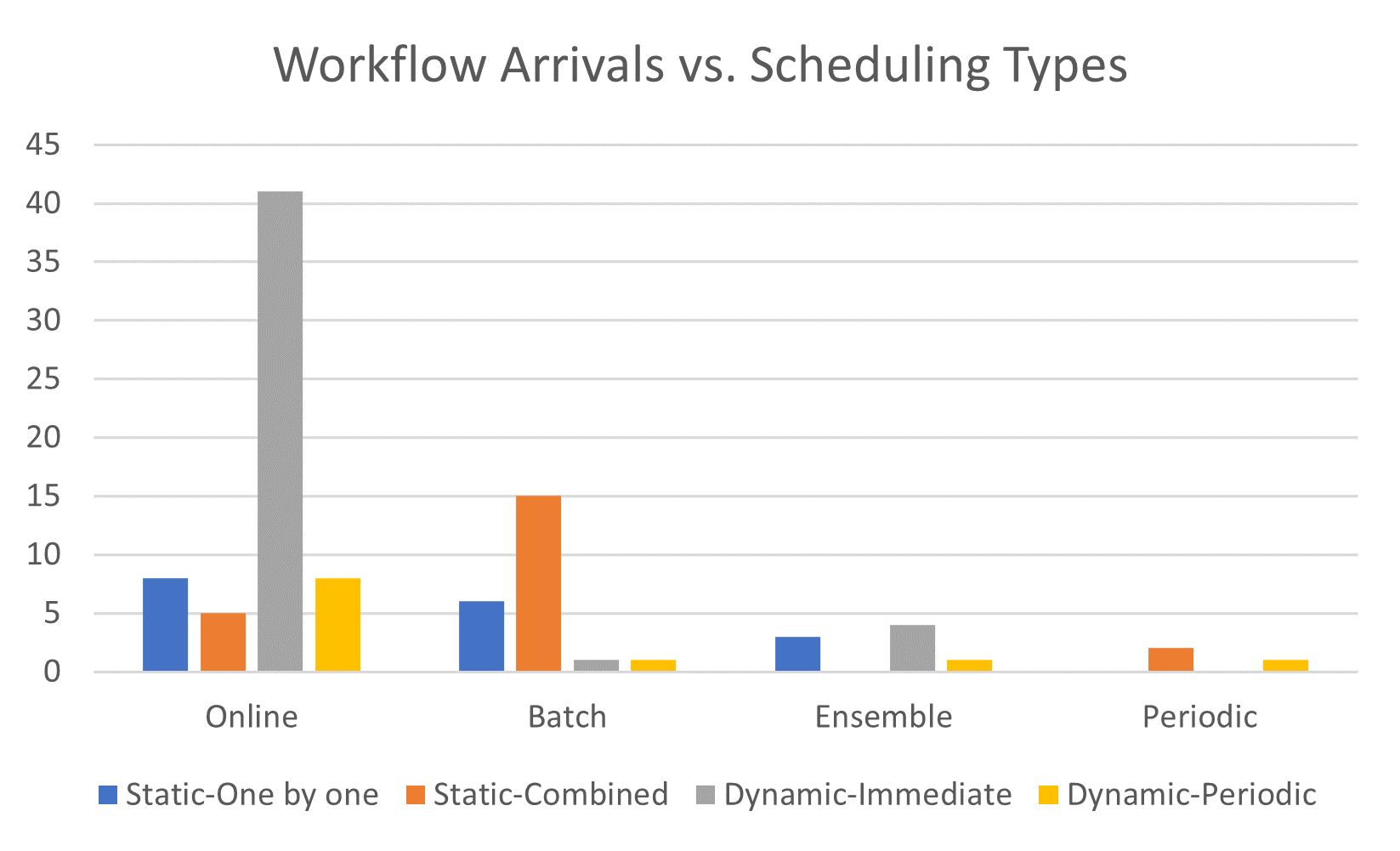}
	%\caption{Workflow Arrivals vs. Scheduling Types}\label{fig:arrType}
	\end{subfigure}
	\caption{Number of Articles by Workflow Arrival Patterns vs. Scheduling Algorithms, and Scheduling Types} \label{fig:arrivalAlgType}
\end{figure}

This section provides a concise overview of the insights derived from previous studies. Figure \ref{fig:arrivalAlgType} illustrates the relationship between workflow arrival patterns and the scheduling algorithms and types reported in the literature. For \textit{online} arrival patterns, a broad range of algorithms has been applied, with \textit{heuristic} and \textit{RL}-based methods being the most prominent. This aligns with the reliance on \textit{dynamic} scheduling to handle the unpredictable and spontaneous nature of online workflows. In contrast, \textit{batch} arrivals are predominantly associated with static scheduling, especially the \textit{combined} type, which leverages prior knowledge of workflow structures and task execution times. Such scenarios also favor \textit{meta-heuristic} algorithms, as they are well-suited to exploring the vast search space created by multiple workflows.
\begin{figure}[t!] % "[t!]" placement specifier just for this example
	\begin{subfigure}{0.5\linewidth}
	\includegraphics[width=\linewidth]{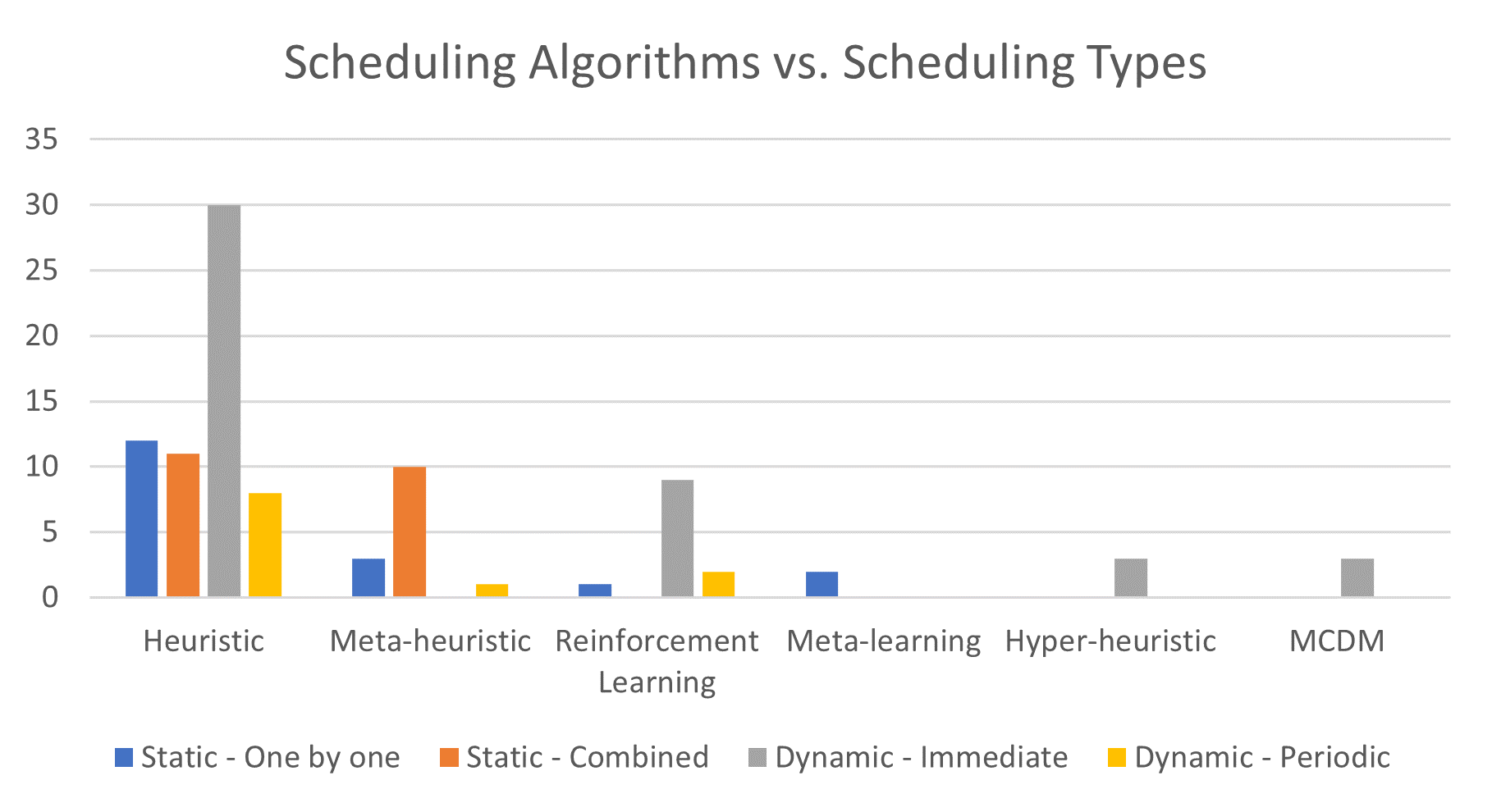}
	\caption{Scheduling Algorithms vs. Scheduling Types}\label{fig:algType}
	\end{subfigure}\hspace*{\fill}
	\begin{subfigure}{0.5\linewidth}
	\includegraphics[width=\linewidth]{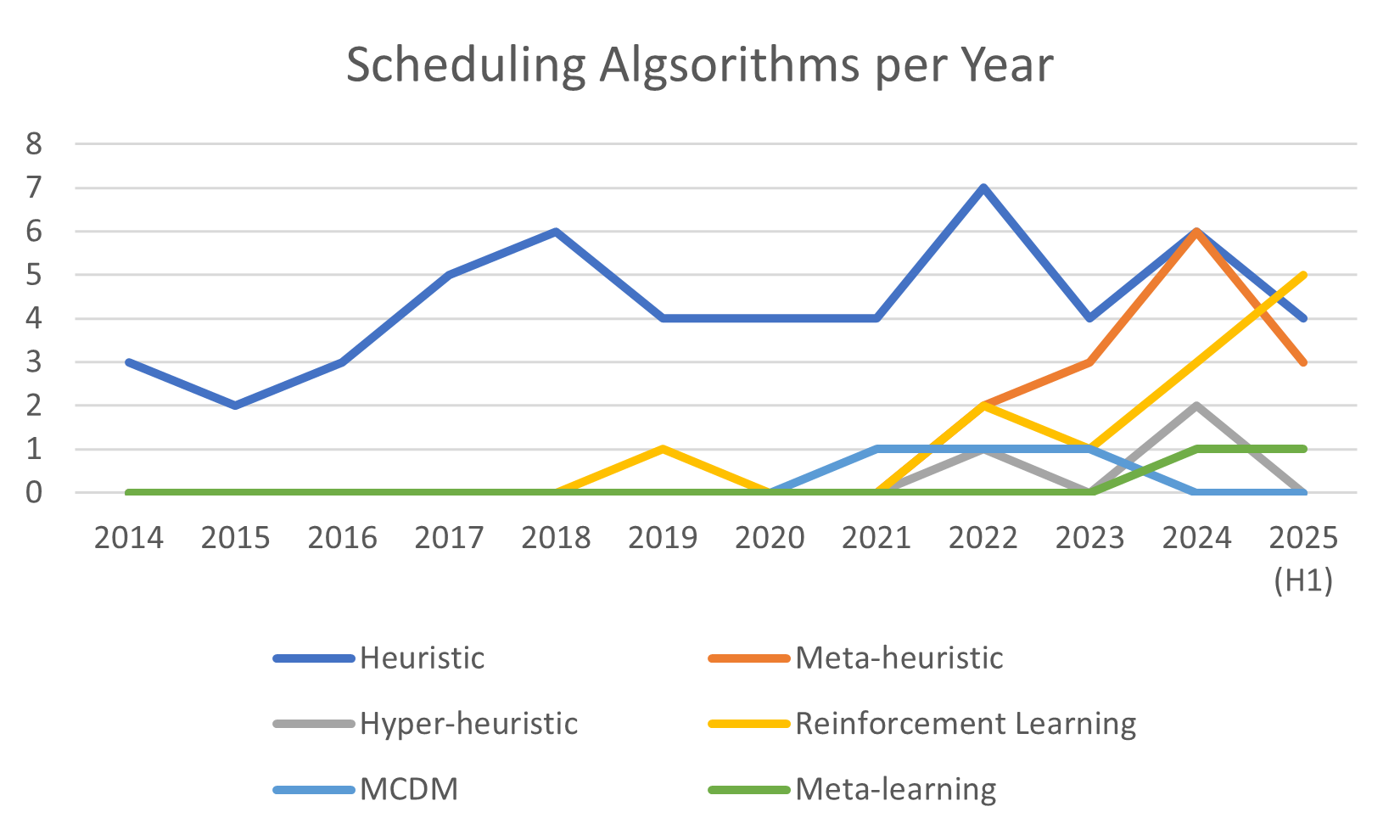}
	\caption{Scheduling Algorithms per Year}\label{fig:algYear}
	\end{subfigure}
	\caption{Number of Articles by Scheduling Algorithms vs. Scheduling Types, and Scheduling Algorithms per Year} \label{fig:AlgTypeYear}
\end{figure}

Figure \ref{fig:algType} summarizes the distribution of scheduling algorithms across different scheduling types. As shown, \textit{heuristic} algorithms have been used in both \textit{static} and \textit{dynamic} scheduling, with a notable emphasis on \textit{dynamic-immediate} strategies. In contrast, \textit{meta-heuristic} algorithms appear exclusively in \textit{static} scheduling, while \textit{RL}-based techniques are predominantly applied in \textit{dynamic} solutions. Figure \ref{fig:algYear} further illustrates the temporal trend in algorithm usage, revealing a growing interest in \textit{RL}, alongside \textit{meta-heuristics}, particularly in recent years—even though the dataset extends only to the first half of 2025.

To conclude, since scheduling algorithms form the core of WaaS brokers and directly impact their overall performance, we provide a comparative evaluation in Table \ref{tab:schAlg}. This table highlights the trade-offs among different scheduling algorithms, including their strengths and weaknesses, their suitability for static or dynamic contexts, and considerations such as execution complexity and scalability.

\begin{table}[t!]
	\centering
	\tiny
	\begin{tabularx}{\textwidth}{|p{1.8cm}||p{2.8cm}|p{2.8cm}|p{2cm}|p{2cm}|}
		\hline
		\textbf{Algorithm} & \textbf{Advantages} & \textbf{Disadvantages} & \textbf{Static/Dynamic} & \textbf{Complexity} \\
		
		\hline
		Heuristic &
		Fast and lightweight, \newline
		Suitable for real-time environments, \newline
		Scales well to large workflows. &
		Typically produces suboptimal schedules, \newline
		Limited adaptability to dynamic changes, \newline
		Lacks generalizability. &
		Suitable for both&
		Very low complexity, \newline
		Typically linear or near-linear \\

		\hline
		Meta-heuristic &
		Good global exploration, \newline
		Effective for complex, multi-objective problems \newline
		Capable of exploring large spaces. &
		High computational cost, \newline
		Slow convergence, \newline
		High parameter sensitivity. &
		Mostly static &
		Moderate to high complexity, \newline
		Scales poorly with tasks/VMs \\
		
		\hline
		Hyper-heuristic &
		Learns from past runs, \newline
		Promotes generality and reusability, \newline
		Combine strengths of multiple heuristics. &
		Long training time, \newline
		Complex design, \newline
		May overfit to training scenarios. &
		Static training, dynamic scheduling &
		Very high during training, \newline
		Scheduling is very fast \\
		
		\hline
		Reinforcement Learning &
		Learns from environment interactions, \newline
		Adapts well to dynamic and uncertain conditions, \newline
		Learns long-term trade-offs. &
		Long training time, \newline
		Reward and feature engineering needed, & 
		Static training, usually dynamic scheduling &
		High training cost, but scheduling is fast and scalable \\
		
		\hline
		MCDM &
		Transparent and interpretable, \newline
		Fast runtime, \newline
		Supports explicit trade-off modeling. &
		Limited scalability to large decision spaces, \newline
		Weight assignment can be subjective, \newline
		Typically used only for VM selection. &
		Usually dynamic. &
		Low complexity. \\
		
		\hline
		Game Theory &
		Models interactions among competing objectives, \newline
		Offers equilibrium-based fairness and stability, \newline
		Models decentralized or competitive scenarios. &
		High mathematical complexity, \newline
		Requires accurate utility modeling. &
		Usually static  &
		High complexity, \newline
		Not scalable beyond few players. \\

		\hline
		Mathematical Optimization &
		Theoretical optimality, \newline
		Transparent formulation, \newline
		Can encode deadlines and resource constraints precisely. &
		Difficult to scale to large instances, \newline
		Solver overhead, \newline
		May not be real-time feasible. &
		Usually static &
		Very high complexity \\
		
		\hline 
\end{tabularx}
\caption{Comparative evaluation of scheduling algorithms.}
\label{tab:schAlg}
\end{table}

\section{Future Directions}
\label{sec:future}
Despite significant advancements in WaaS brokers, numerous open challenges and research opportunities remain. The rise of artificial intelligence (AI) and large language models (LLMs) opens new possibilities for transforming the design and operation of WaaS brokers through intelligent decision-making, predictive resource management, and automated negotiation, potentially rendering many traditional methods obsolete. Emerging technologies, including serverless and quantum computing, further expand opportunities to enhance broker performance and flexibility. Additionally, underexplored areas such as dynamic negotiation strategies and fault-tolerant mechanisms provide promising directions for improving user satisfaction and overall system robustness. This section highlights these opportunities, aiming to guide researchers toward the most impactful avenues for advancing WaaS broker design and implementation.

\subsection{AI-Driven WaaS Brokers}
\label{sec:AI-driven}
Recent advances in AI, particularly in generative LLMs and DRL, are transforming the landscape of cloud computing. WaaS brokers enhanced with these technologies could harness them throughout the workflow lifecycle, potentially redefining both their design and performance.

DRL enables dynamic decision-making in uncertain and continuously changing cloud environments. By learning optimal policies for task scheduling and resource provisioning through direct interaction with the environment, DRL-based brokers can adapt to workload fluctuations, heterogeneous resources, and shifting performance goals. This adaptability makes scheduling decisions more efficient and resilient. Figure \ref{fig:algYear} illustrates that reinforcement learning has already received strong research attention for workflow scheduling in recent years, and its potential for further applications remains significant.

Meta-RL \cite{beck2025tutorial} extends these capabilities by enabling brokers to quickly adapt to new workflow scenarios with minimal experience. Instead of tailoring policies only for specific environments, meta-RL learns a meta-policy that generalizes across many scenarios. In the context of multiple workflows, this means the scheduler can leverage knowledge gained from previously encountered workflow structures, arrival patterns, and resource conditions to make faster and more effective decisions for unseen workloads. During training, brokers are exposed to synthetic but diverse situations—such as varying workflow DAGs, fluctuating resource availability, or dynamic pricing. As a result, the meta-policy learns general adaptation strategies rather than memorizing solutions for specific cases, making it especially suitable for multi-tenant WaaS environments where workflow types, deadlines, and system conditions can change unpredictably.

LLMs bring a different but complementary strength. They introduce adaptive, context-aware reasoning into scheduling and provisioning decisions, which were traditionally driven by fixed rules or heuristics. An LLM-enhanced broker can parse high-level workflow descriptions, SLAs, and performance objectives expressed in natural language, translating them into executable scheduling and provisioning plans. It can analyze task dependencies, predict execution times from logs, and choose suitable resource types, whether serverless or traditional VMs, and pricing models. With access to real-time monitoring data, LLMs can also detect SLA violations or anomalies, identify root causes, and recommend or directly trigger corrective actions such as scaling resources or rescheduling tasks. Beyond automation, LLMs can serve as a natural interface between users and brokers, allowing administrators to specify goals in plain language—for example: "allocate spot instances to flexible tasks"—which the broker then translates into scheduling directives.

%Although currently there are no previous study in using LLMs for resource provisioning and scheduling of scientific workflows, there are a few research in applying LLMs to job scheduling in cloud environments with most works currently appearing as non-peer-reviewed preprints. Jadhav et al. \cite{jadhav2025evaluating} introduced ReAct (Reason+Action), an LLM-based scheduler for multi-objective job scheduling in HPC environments. Wu et al. \cite{wu2025learning} proposed a hierarchical language agent framework that automatically generates context-aware scheduling policies using LLM reasoning capabilities to schedule VMs on physical machines in cloud computing. In another notable example, Akbari et al. \cite{akbari2025intent} developed IntentContinuum, a novel framework for intent-driven resource management across the compute continuum of edge and cloud resources. In this framework, \textit{intents} represent high-level user goals, such as minimizing latency or optimizing energy consumption. The LLM-based decision maker processes cluster, network, and monitoring information in a machine-readable form. If a deviation from the expected performance (intent) is detected, it identifies the root cause using a structured violation template and then applies a recommended action template to propose targeted corrective measures. 

Research in this area is still nascent. While no studies currently apply LLMs directly to workflow scheduling, a few early works explore their use in related contexts. Jadhav et al. \cite{jadhav2025evaluating} introduced ReAct (Reason+Action), an LLM-based scheduler for multi-objective job scheduling in HPC environments. Wu et al. \cite{wu2025learning} proposed a hierarchical language-agent framework that generates context-aware VM scheduling policies using LLM reasoning capabilities to schedule VMs on physical machines. Similarly, Akbari et al. \cite{akbari2025intent} introduced IntentContinuum, an LLM-based framework for intent-driven resource management across edge–cloud systems. User goals (e.g., minimizing latency or energy use) are expressed as intents, which the framework monitors against system performance. When violations occur, the LLM identifies root causes and suggests corrective actions through structured templates.

In many cases, the general reasoning capabilities of state-of-the-art LLMs are sufficient to generate effective scheduling decisions when these inputs are clearly provided in natural language or structured formats. However, for scientific workflows—especially multiple concurrent workflows—scaling such approaches requires grounding LLMs with domain knowledge. Retrieval-Augmented Generation (RAG) can improve accuracy by incorporating historical workflow traces or performance data, while fine-tuning can help internalize specialized scheduling strategies. Together, these techniques improve efficiency, consistency, and adaptability, enabling brokers to balance cost, performance, and deadline requirements in highly dynamic environments.

Another promising direction is the integration of LLMs and RL for scheduling and provisioning in WaaS brokers. In this setup, the LLM functions as a high-level strategist, generating candidate scheduling and provisioning plans by reasoning over workflow deadlines, dependencies, available resources, and pricing schemes. These plans are then passed to an RL agent, which interacts with the environment to evaluate their effectiveness against performance metrics such as makespan, cost, energy efficiency, and SLA compliance. Through continuous feedback, the RL agent refines and adapts the strategies, ensuring they are well-tuned to real-world dynamics. This synergy combines the broad reasoning and contextual awareness of LLMs with the optimization power of RL, enabling brokers to be both innovative in planning and robust in execution. While this approach has not yet been applied to workflow scheduling specifically, similar LLM–RL frameworks have shown promise for independent task scheduling \cite{tang2024llm, krishnamurthy2025large}.

\subsection{Toward Intelligent Negotiation and Autonomous Fault Tolerance}
\label{sec:towardNegotiation}
Two dimensions that remain surprisingly underexplored in WaaS broker research are negotiation/pricing mechanisms and fault tolerance strategies. Yet, these components hold the potential to redefine the role of the broker from reactive middleware into autonomous, market-aware, and self-healing ecosystems.

%On the negotiation side, current practice still relies on static pricing schemes and rigid SLAs, which cannot fully capture the dynamic realities of multi-tenant environments. Future WaaS brokers could instead embrace predictive, multi-objective negotiation engines, continuously adjusting SLAs in response to input workload characteristics, resource availabilty and cost. In multi-criteria optimization scenarios, brokers might identify Pareto-optimal solutions and present them to users as menus of possibilities—for example, a set of non-inferior QoS tiers, each linked to distinct price points. Users could then make informed choices, such as selecting higher-cost, higher-reliability execution paths for sensitive workflows. Beyond this, pricing itself demands a paradigm shift: rather than cost-based formulas, brokers could explore value-driven and market-based pricing inspired by economics \cite{wu2019cloud}. Such policies could incentivize workload submissions during off-peak hours when reserved instances are idle, or encourage users with flexible deadlines to unlock savings by enabling the broker to exploit spot resources. In this vision, the broker evolves into a market-aware strategist, aligning the incentives of users and providers while maximizing efficiency, fairness, and sustainability.
On the negotiation side, most existing approaches still depend on static pricing models and rigid SLAs, which do not capture the dynamic nature of multi-tenant cloud environments. Future brokers could instead employ predictive, multi-objective negotiation engines that continuously adapt SLAs to workload demands, resource availability, and market conditions. By computing Pareto-optimal solutions in multi-criteria optimization settings, brokers could present users with structured QoS–price trade-off options (e.g., higher reliability at a higher cost). Furthermore, pricing strategies should evolve from simple cost-based formulas to value-driven, market-oriented models \cite{wu2019cloud}, incentivizing behaviors such as off-peak submissions when reserved resources are underutilized or flexible deadlines that take advantage of spot instances. In this vision, the broker acts as a market strategist, harmonizing user and provider incentives while fostering both efficiency and fairness.

LLMs reinforce this vision by enabling adaptive, human-like negotiation. Unlike traditional rule-based, LLMs can engage in conversational bargaining with users, interpreting natural concerns such as "the cost is too high", and proposing trade-offs such as offering a reduced price in exchange for a longer completion time. LLMs are also able to analyze market conditions, compare pricing models, and take SLA constraints into account when predicting the likely cost of execution and generating adaptive offers or counteroffers. By retrieving historical negotiation outcomes and pricing data via RAG, brokers can further ground their strategies in real-world patterns, making the negotiation process more efficient, transparent, and user-friendly. Recent work on negotiation between humans and software agents using LLMs \cite{bianchi2024how, schneider2025negotiating, vahidov12025using} highlights promising directions that could be directly adopted in WaaS broker negotiation frameworks.

Fault tolerance is another area where a fundamental rethinking is necessary. At present, brokers largely depend on reactive strategies like retries or failovers. Future designs should combine predictive modeling, hybrid strategies, and proactive resilience. Depending on the context, replication could leverage idle reserved instances, re-execution could handle short tasks, and checkpointing could protect long-running jobs. DRL could be leveraged to identify tasks that are critical or highly fault-prone, and to determine whether they should be duplicated or checkpointed to minimize the impact of failures. Crucially, this layer could be intertwined with negotiation and pricing, enabling users to select reliability levels as a priced option—turning fault tolerance itself into a negotiable, market-integrated feature. More radically, by combining \textit{digital twins} \cite{iranshahi2025digital} of datacenters with survival analysis and deep learning, brokers could simulate cascading failure scenarios in real time, migrating or replicating workflow components before bottlenecks or SLA violations occur. Over time, such systems would refine their resilience through continual feedback loops, effectively learning \textit{what-if} strategies for complex multi-tenant environments. 

LLMs can also significantly enhance monitoring and fault-tolerance functions by converting raw telemetry into actionable insights. Unlike traditional monitoring systems that rely on rule-based anomaly detection, LLMs can analyze heterogeneous inputs such as logs, metrics, error codes, and workload traces to identify anomalies, determine the root cause of performance degradations, and provide human-readable diagnoses. In the context of fault tolerance, LLMs can reason about appropriate recovery strategies—such as retrying transient failures, migrating tasks to alternative resources, duplicating critical tasks, or triggering checkpoints—based on workflow topology, task criticality, and SLA constraints. Moreover, LLMs can learn or retrieve patterns from historical execution data to guide proactive decisions; for example, "for DAGs with long critical paths, prioritize checkpointing of bottleneck tasks." By integrating such knowledge through RAG, LLMs can ground their reasoning in domain-specific logs, past workflow traces, and known fault scenarios, ensuring more accurate and context-aware recommendations, such as avoiding unreliable spot instances for deadline-sensitive workflows. 

Traditionally, DevOps engineers handled fault detection, localization, and mitigation. The concept of AI for IT Operations (AIOps) has been proposed as a means of automating these processes by applying machine learning to the large volumes of monitoring and infrastructure data, thereby enabling automatic detection, diagnosis, and resolution of system issues with minimal human involvement \cite{shetty2024building}. Recent advances in AIOps have introduced LLM-based agents that exploit natural language processing to interpret diverse input data and logs, identify causes of faults, and even generate scripts that invoke external tools for automated remediation \cite{shetty2024building, yang2025cloud}. Zhang et al. \cite{zhang2025survey} provide a recent survey of LLM-based AIOps systems. Adapting such techniques to WaaS brokers could reduce operational overhead while improving robustness and resilience.

\subsection{WaaS Brokers as Cloud Economists}
\label{sec:economist}
Although some studies have explored combining pricing models (on-demand, reserved, and spot) for WaaS brokers, more sophisticated approaches are still needed. A promising direction is to reconceptualize brokers not just as technical mediators between users and CSPs, but as intelligent \textit{cloud economists}. By leveraging multiple pricing models, brokers could act as market agents that strategically navigate uncertainty and volatility. Instead of treating pricing models as static choices, future brokers could dynamically orchestrate them based on long-term workload predictions and short-term market signals. For example, they might proactively acquire reserved capacity for recurring workflow patterns, exploit spot markets for opportunistic savings, and rely on on-demand instances to meet deadlines. In this way, brokers function as self-optimizing portfolios of cloud resources, balancing risk, cost, and performance much like financial traders manage assets in dynamic markets.

This vision can be realized through integrating LLMs and RL into a unified decision-making loop. LLMs, with their ability to interpret workflow descriptions, pricing policies, and historical traces, can serve as \textit{strategists} that map user requirements to high-level provisioning strategies. RL agents, in turn, act as \textit{executors}, refining these strategies under uncertainty by adapting to workload variability and market fluctuations. With RAG grounding decisions in past negotiation outcomes and cost histories, brokers could evolve into proactive, self-learning agents that bargain with users, forecast demand, and arbitrage across pricing models. WaaS brokers, therefore, evolve beyond simple workflow managers, emerging as autonomous, market-aware entities—cloud economists that redefine how scientific workflows are priced, provisioned, and executed at scale.

\subsection{Federated Brokers for Cross-Cloud and Data-Sensitive Workflows}
As workflows increasingly span multiple administrative domains or involve sensitive data in domains such as healthcare, finance, and genomics, WaaS brokers must ensure privacy, data locality, and regulatory compliance. Traditional centralized brokers collect the full workflow DAG and raw data at a single control point, which simplifies orchestration but exposes sensitive inputs and may violate legal or organizational constraints. Federated brokers, in contrast, adopt a decentralized architecture where each participating domain or CSP retains control of its data and tasks, contributing only high-level metadata, encrypted updates, or aggregated statistics to the global decision process. Each cloud region or administrative domain hosts a local broker agent, and these agents coordinate—optionally guided by a central WaaS broker—to orchestrate workflows while enforcing local policies.

In such a setting, federated brokers must perform task graph partitioning under policy-based placement constraints. Tasks involving sensitive data can remain on-premise or within a specific trusted cloud, while other tasks can be flexibly scheduled to public clouds for efficiency or cost reduction. To coordinate these distributed decisions, federated brokers can employ federated learning to train predictive models for task runtimes across domains, secure multi-party computation or homomorphic encryption for joint optimization of provisioning, and differential privacy to protect workflow-level analytics. This allows brokers to orchestrate end-to-end workflows across heterogeneous clouds while guaranteeing privacy, compliance, and performance scalability.

\subsection{Hybrid Serverless/Serverful Provisioning}
\label{sec:hybridServerless}

The rise of serverless computing and the FaaS model introduces new opportunities for WaaS brokers \cite{wen2023rise}. In this model, users and brokers no longer manage resources or scheduling directly; instead, the CSP handles provisioning, scaling, and function orchestration. FaaS also offers fine-grained, utilization-based pricing, charging only for actual consumption and thus eliminating idle-resource overheads. While several studies have explored applying FaaS to scientific workflows \cite{malawski2020serverless}, its adoption remains in the early stages. Existing work mostly targets single workflows, with frameworks such as StepConf \cite{wen2022stepconf}, Wukong \cite{carver2020wukong}, SWEEP \cite{john2019sweep}, and StarShip \cite{roy2024starship}. Other research has investigated multi-cloud scheduling across multiple CSPs, including xAFCL \cite{ristov2023xafcl}, XFaaS \cite{khochare2023xfaas}, SwitchFlow \cite{chen2024switchflow}, and FaaSt \cite{ristov2022faast}.

Despite these advances, several limitations hinder workflow execution on pure serverless platforms \cite{roy2022mashup}. First, the separation of computation and storage, intended to enable auto-scaling, results in stateless functions that cannot retain intermediate states. This creates inefficiencies for workflows requiring data exchange between tasks, forcing reliance on external storage, which is slower than direct task-to-task transfers \cite{li2023serverless}. Second, serverless environments are optimized for lightweight, short-running tasks, with execution time limits (e.g., 300s in AWS Lambda, 60s in Google Cloud Functions), making them unsuitable for long-running long-running tasks. Finally, the \textit{scale-to-zero} mechanism de-provisions idle function containers after a period of inactivity, leading to \textit{cold start} delays that particularly affect short tasks.

To overcome these issues, hybrid provisioning strategies have been proposed, combining serverful and serverless platforms \cite{malawski2020serverless}. In such environments, the weaknesses of each model are offset by the strengths of the other: VMs are suited for long-running or sequential tasks, while functions excel at lightweight, parallel workloads. Local VM storage can facilitate large data exchanges between tasks by eliminating transfer overheads, while serverless functions reduce underutilization and improve elasticity. Serverless services can also be paired with various VM pricing models to lower both cost and makespan. DRL and LLM-based decision engines may further optimize task placement by deciding whether a task should run as a function or on a VM based on execution time, data size, or deadlines, and by selecting the most cost-effective pricing model.

Several studies have investigated hybrid execution, though most focus only on single workflows. Notable examples include Mashup \cite{roy2022mashup}, Lithops \cite{eizaguirre2024serverful}, DEWE v3 \cite{jiang2017serverless}, HyperFlow on AWS Lambda \cite{balis2020cloud}, and the Budget-Constrained Workflow Scheduling (BCWS) algorithm \cite{zhang2024scheduling}. The only existing work that considers multiple workflows in a hybrid environment is by Sun et al. \cite{57}, which is covered in this SMS. Overall, the integration of serverless FaaS with serverful VMs under diverse pricing models offers a highly promising direction for WaaS brokers. Achieving this vision calls for new provisioning and scheduling algorithms capable of fully exploiting hybrid platforms to minimize costs and improve performance in scientific workflow execution.

\subsection{Hybrid Quantum/Classical WaaS Brokers}
\label{sec:hybridQuantum}

As discussed earlier, quantum computing remains an emerging, rapidly evolving technology, with current systems offering only limited capabilities. Consequently, hybrid HPC/quantum environments—and the corresponding hybrid WMSs—are still in the research and development phase. The architectural and behavioral differences between quantum and classical computers make existing WMSs and their provisioning and scheduling algorithms inadequate for hybrid scientific workflows. Cranganore et al. \cite{cranganore2024paving} highlight the main challenges of hybrid workflow execution and propose a high-level architectural design for a hybrid WMS built on Pegasus, detailing essential software components, their roles, and the system-level requirements.

A first major challenge is the heterogeneity of quantum hardware technologies—such as superconducting qubits, trapped ions, and optical/photonics—which differ in number of qubits, supported quantum gates, and coherence times \cite{beck2024integrating}. These variations may render certain algorithms inefficient or even infeasible on specific hardware. Moreover, no standard set of performance descriptors yet exists. While metrics like IBM’s Quantum Volume (QV) and Circuit-Layer Operations per Second (CLOPS) have been introduced, they remain evolving and non-standardized. To cope with this heterogeneity, a hybrid WaaS broker must provide interfaces to multiple quantum backends and support transpilation of high-level circuits into hardware-specific representations—assuming execution is feasible. This requires a flexible programming interface that abstracts hardware details, ensuring portability and adaptability across quantum platforms \cite{shehata2026bridging}.

A second challenge is developing accurate performance models for predicting execution time, error rates, and other key metrics on specific quantum hardware. Such models are critical for broker schedulers to make informed allocation decisions, yet constructing generalizable models across quantum platforms remains unresolved \cite{cranganore2024paving}. Other challenges include hybrid workflow optimization—through techniques like gate-level minimization to reduce circuit depth—and error mitigation in noisy intermediate-scale quantum (NISQ) devices, where noise and decoherence still hampers reliability.

Despite these challenges, some researchers have proposed initial frameworks designed to define and execute single hybrid workflows in quantum/classical environments. Notably, Weder et al. introduced MODULO \cite{weder2021modulo, weder2023provenance}, a hybrid WMS with components for workflow generation, optimization, packaging, deployment, and monitoring. While their evaluation targets a single workflow, the design could extend to multiple workflows. Other efforts include Pilot-Quantum \cite{pradeep2025pilot}, XFaaSQ—a FaaS-based hybrid workflow engine \cite{jha2025choreography}, and QCCP (Classical-Quantum Collaborative Computing) \cite{du2025qccp}.

The only current WMS explicitly supporting multiple hybrid workflows is Qonductor \cite{giortamis2024orchestrating}, still a preprint in mid-2025. Its modular architecture consists of a client, a leader, and one or more workers. The leader node houses the resource estimator, hybrid scheduler, and job manager. The estimator generates alternative resource plans based on workflow fidelity and estimated execution time; clients select one plan, which is then executed by the job manager. The job manager coordinates the execution of ready jobs by invoking the hybrid scheduler to allocate appropriate resources. The hybrid scheduler applies a simple filtering-scoring heuristic for classical jobs, and a three-stage process for quantum jobs—preprocessing, optimization, and selection. During optimization, NSGA-II produces a Pareto front balancing completion time and fidelity, and then an MCDM method selects the best option. Qonductor also includes monitoring and fault-tolerance mechanisms to improve reliability. Importantly, its evaluation demonstrates multiple workflow scheduling and execution—an advance over existing work.

In parallel, researchers at Oak Ridge National Laboratory (ORNL) have developed a robust framework for embedding QC into classical HPC environments \cite{beck2024integrating, shehata2026bridging}. It features a unified resource manager orchestrating both quantum and classical resources, along with a versatile programming interface that abstracts hardware differences. Their proposed architecture is notably rich and addresses a wide range of requirements for a cohesive hybrid environment. It introduces key abstractions for task decomposition and coordination, enabling efficient interaction between quantum and classical tasks. Such designs can be adopted by WMSs like Pilot-Quantum to facilitate hybrid workflows.

Looking ahead, research should focus on developing hybrid quantum/classical WaaS brokers with provisioning and scheduling algorithms tailored for quantum workloads. These brokers must consider quantum-specific challenges, including diverse performance descriptors and hardware-dependent performance models, to achieve efficient, scalable hybrid workflow execution.

\section{Conclusion}
\label{sec:conclusion}
This study conducts an SMS on Workflow as a Service (WaaS) brokers in cloud environments, covering the period from 2009 to July 2025. The SMS identified 87 relevant articles and 49 venues using a comprehensive search strategy. Based on the analysis of the extracted articles, a taxonomy has been proposed by considering the key components of WaaS brokers, namely \textit{negotiation and pricing}, \textit{task scheduler}, \textit{resource provisioner}, and \textit{monitoring and fault tolerance}. All articles have been classified and reviewed according to this taxonomy.  

The results reveal that WaaS brokers have attracted increasing attention, particularly in the past three years. While notable progress has been achieved in provisioning and scheduling, other components remain underexplored, offering significant opportunities for future research. Looking ahead, several promising directions can further expand the role of WaaS brokers. Advances in AI, particularly in DRL and LLMs, can enable brokers to act as intelligent, autonomous agents capable of adaptive negotiation, dynamic SLA management, and predictive scheduling. The convergence of serverful and serverless paradigms opens new opportunities for hybrid resource provisioning strategies that can balance performance, cost, and elasticity. Moreover, the rise of quantum computing suggests the need for brokers that can orchestrate hybrid quantum/classical workflows, requiring novel provisioning and scheduling algorithms tailored to quantum workloads. Together, these directions highlight how future WaaS brokers can evolve from workflow managers into intelligent, market-aware orchestrators at the heart of next-generation cloud ecosystems.

\bibliographystyle{elsarticle-num.bst}
\bibliography{References}

\appendix
\newpage
\section{Industrial WaaS Brokers}
In this section, we examine the real-world adoption and practical deployment of WaaS brokers. To manage the composition, planning, orchestration, and automation of scientific workflows efficiently, a wide range of Workflow Management Systems (WMSs) have been developed, many of which can serve as the foundation for WaaS brokers. Suter et al. \cite{suter2025terminology} surveyed the properties and specifications of 23 widely used WMSs that are part of the Workflows Community Initiative (WCI). Most of these systems support workflow execution on local resources, HPC clusters (via resource managers such as LSF, PBS, Slurm, Torque, HTCondor, and Grid Engine), and cloud infrastructures, particularly the three leading providers—AWS, GCP, and Azure. While not all WMSs possess the full set of capabilities required to qualify as WaaS brokers, several come close, and others have been adapted as the basis for practical broker implementations. We begin with open-source and research prototypes, followed by an overview of commercial solutions, including broker offerings from the major cloud providers. Table \ref{tab:waas_brokers} provides a comparison of selected real-world WaaS brokers.

\begin{figure*}[t] % "[t!]" placement specifier just for this example
	\includegraphics[width=\linewidth]{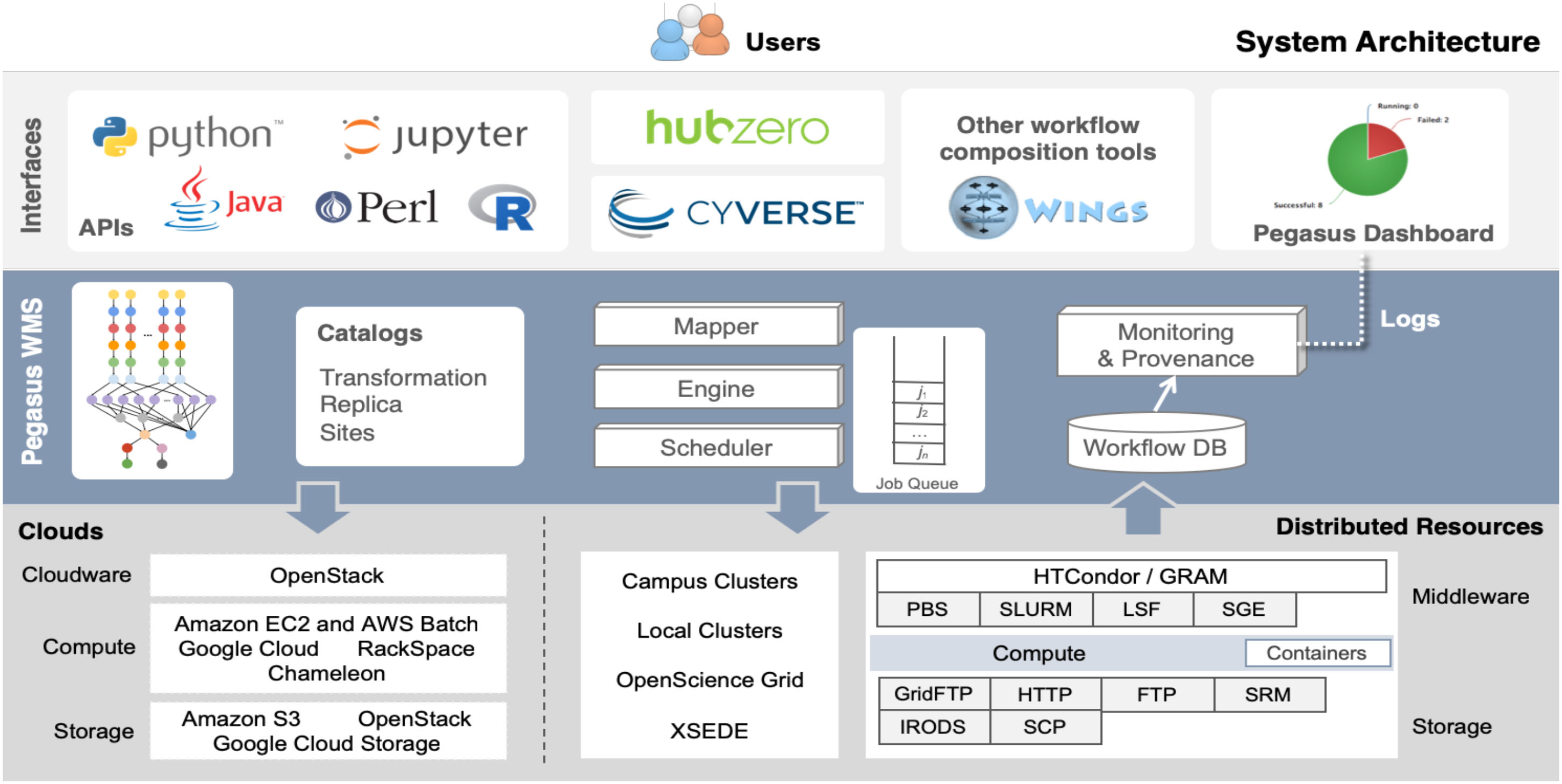}
	\caption{The architecture of Pegasus workflow management system \cite{deelman2019evolution}}
	\label{fig:pegasus}
\end{figure*}

One of the most known WMS is Pegasus \cite{deelman2019evolution} which has been used in many scientific domains including astronomy, bioinformatics, earthquake science , gravitational wave physics, and others. Figure \ref{fig:pegasus} illustrates the architecture of Pegasus. As can be seen, Pegasus has four main components: \textit{Mapper}, \textit{Scheduler}, \textit{Engine}, and \textit{Monitoring and Provenance}. The \textit{Mapper} generates an executable workflow based on an abstract workflow representation. The executable workflow has been sent to a workflow \textit{Engine} which is responsible for submitting the tasks to the resources according to their dependencies. Pegasus supports \textit{local workflow execution engine} for executing tasks on a local cluster, and \textit{remote workflow execution engine} for executing tasks on remote resources such as cloud. The execution engine follows a \textit{just-in-time} planning, so once a task becomes ready, it calles the \textit{Scheduler}. \textit{Scheduler} manages the execution of workflows' tasks on local and remote resources, and provides task-level reliability. To decrease scheduling overhead, Pegasus clusters small \textit{tasks} into larger, more compute-intensive \textit{jobs}. Furthermore, it uses \textit{workflow partitioning} algorithms to divide the workflow graph into smaller portions, and schedules a single workflow partition at a time. Finally, the \textit{Monitoring} component, monitors the progress of the workflow, and populates the jobs and task logs into a database. Pegasus can be integrated with AWS Batch and Google Cloud Batch to efficiently provision resources for executing workflows. 

\begin{table}[t]
\centering
\tiny
\begin{tabularx}{\textwidth}{|l|c|X|X|X|X|}
\hline
\textbf{Broker} & \textbf{Comm.} & \textbf{Pricing} & \makecell{\textbf{Execution}\\\textbf{Environment}} & \makecell{\textbf{Workflow}\\\textbf{Language(s)}} & \makecell{\textbf{Workflow}\\\textbf{Engine}} \\
\hline
Pegasus & No & Free (open-source) & Local, HPC, Cloud (AWS, GCP, Azure) & DAX (XML), Python, CWL (limited) & Pegasus WMS \\
\hline
Airavata (Apache) & No & Free (self-hosted) & HPC, Cloud (via wrappers) & CWL, JSON API & Airavata Orchestrator \\
\hline
REANA (CERN) & No & Free (self-hosted) & Local (Docker), Kubernetes, Cloud (via K8s) & CWL, Yadage, Snakemake, Serial & REANA Engine (pluggable) \\
\hline
Tibanna & No & Free; AWS costs apply & AWS (EC2 + Lambda + Step Functions) & CWL, WDL & Cromwell / cwltool \\
\hline
Terra & Partial & Free UI; GCP billed & GCP & WDL, CWL & Cromwell \\
\hline
Arvados & No & Free (self-hosted) & Local, HPC, Cloud (AWS, GCP) & CWL & Crunch Engine \\
\hline
Seqera Platform Cloud & Yes & Free~(basic)~, Paid~(Pro/Ent) & HPC, Cloud (AWS, Azure, GCP)& Nextflow & Nextflow Engine \\
\hline
MWAA & Yes & Baseline cost + Pay-Per-Use & AWS & Python & Apache Airflow \\
\hline
Vertex AI Pipelines & Yes & Pay-Per-Use & GCP & Python & Apache Airflow \\
\hline
Azure Data Factory & Yes & Pay-Per-Use & Azure & Visual interface + Python & Proprietary + Apache Airflow \\
\hline
DNAnexus & Yes & Enterprise pricing & AWS (native), Azure, GCP (via partners) & CWL, WDL, Nextflow, Python & Proprietary + Cromwell/Nextflow \\
\hline
Seven Bridges & Yes & Enterprise pricing & AWS (native), GCP, Azure (some support) & CWL, WDL & Rabix Executor \\
\hline
\end{tabularx}
\caption{Comparison of selected open-source and commercial WaaS brokers.}
\label{tab:waas_brokers}
\end{table}

Apache Airavata \cite{Airavata} is another open-source scientific WMS, which helps researchers and developers build and operate Science Gateways — web-based platforms for running computational jobs, simulations, and data analysis workflows on HPCs, grids, and clouds. It act as a WaaS broker in some universities and institution such as SEAGrid Science Gateway and SimVascular Gateway. It has an internal XML-based format for describing DAG-like workflows, and can execute workflows on local, HPC clusters, and cloud resources, specifically on AWS EC2 instances and OpenStack platform. However, it does not support SLA/cost-based brokering and pricing plans for the users. Another well-known WaaS broker is REANA (REproducible ANAlysis platform) \cite{Reana}, developed by CERN. It's a modern, cloud-native system that enables researchers to run, share, and reproduce complex scientific workflows, especially in high energy physics and data-intensive research. It supports different workflow languages such as Common Workflow Language (CWL) and Snakemake, and can execute workflows on any Kubernetes cluster such as a local cluster, along with AWS Elastic Kubernetes Service (EKS), Google Kubernetes Engine(GKE), and Azure Kubernetes Service (AKS). 

There also some special purpose WaaS brokers. Tibanna \cite{lee2019tibanna} is an open-source WMS, which is designed to run bioinformatics workflows on AWS, using Lambda, Step Functions and EC2. It accepts workflows in Workflow Description Language (WDL), and CWL. It can also run Snakemake \cite{molder2021sustainable} workflows, which is another popular python-based WMS, on AWS resources. Snakemake uses an extension of Python language to define workflows in terms of rules that define how to create output files from input files. Dependencies between the rules are determined automatically, creating a DAG of jobs.  Snakemake also has different plugins to run workflows on Azure Batch and Google Cloud. Terra \cite{Terra} is another cloud-based platform for biomedical data analysis. It uses Cromwell WMS \cite{Cromwell} to execute genomics and bioinformatics workflows on Google Cloud. It has a pay-as-you-go pricing model for the GCP used resources. Cromwell \cite{Cromwell} is a popular WMS which receives workflows using WDL and executes them on local, HPC clusters and cloud resources (Google Cloud Batch and AWS Batch). Arvados \cite{Arvados} is another open source platform, act as a WaaS broker for bioinformatic workflows. It supports CWL for defining workflows and can executes workflows on on-premise HPC clusters, as well as AWS, GCP and Azure cloud platforms. It also can use preemptable (spot) instances to decrease the execution costs. 

There are also some commercial WaaS brokers. Seqera Platform \cite{SeqeraPlatform} is a web-based platform for executing, managing, visualizing, and monitoring Nextflow pipelines. Nextflow \cite{Nextflow} is another popular WMS for scientific workflows, which receives the users' workflows in the form of a \textit{pipeline}, defined by a Domain Specific Language (DSL) script which consists of \textit{processes} (units of work) connected by \textit{channels} (data streams between processes). At runtime, Nextflow parses the DSL to build a DAG representing tasks and their data dependencies. It schedules tasks when becomes ready (dynamic scheduling), on different \textit{executors}, including local, HPC schedulers, Kubernetes, or Cloud batch services (AWS Batch, Azure Batch, GCP Batch). Using Seqera Platform, users can easily track the progress of their pipelines, monitor resource usage, and manage access control. Then, Seqera introduced Seqera Platform Cloud \cite{SeqeraCloud}, which is a fully managed platform for running scientific workflows on cloud resources. It can be considered as a real WaaS broker for researchers, offers three main pricing tiers: Cloud Basic (free), Cloud Pro (paid), and Enterprise. 

All three major cloud providers—AWS, GCP, and Microsoft Azure—offer workflow orchestrators for serverless environments: AWS Step Functions \cite{StepFunctions}, Azure Durable Functions \cite{DurableFunctions}, and Google Cloud Workflows \cite{GCWorkflows}. However, these services are generally unsuitable for large-scale, data-intensive, or HPC-style scientific workflows due to limitations such as the lack of native data staging and provenance support, restricted scalability for large task graphs, and runtime limits per task (e.g., 15 minutes in AWS). Despite these limitations, they can be integrated with services like AWS Batch \cite{AWSBatch}, a fully managed batch computing service that supports containerized workloads at scale. AWS Batch includes a cloud-native job scheduler, elastic provisioning of on-demand and spot instances, and built-in monitoring and logging. It can be used in conjunction with popular WMSs such as Pegasus, Luigi, Nextflow, Metaflow, Apache Airflow, and Step Functions to execute scientific workflows. However, it supports only containerized workloads, has a relatively complex setup, and requires manual data staging via services like Amazon S3 or EFS. Azure Batch \cite{AzureBatch} and Google Cloud Batch \cite{GCBatch} serve as Microsoft’s and Google’s counterparts to AWS Batch. These services also support integration with WMSs like Nextflow for scientific workflows. Among them, Google Cloud Batch is the most user-friendly, as it eliminates the need for users to define compute environments or manage resource pools, and it provides native integration with Cloud Storage, removing the need for manual data staging scripts.

Apache Airflow \cite{Airflow} is an open-source WMS used to define, schedule, and monitor workflows. Airflow workflows are represented as DAGs, written entirely in Python—an approach commonly referred to as workflows as code. Amazon Managed Workflows for Apache Airflow (MWAA) \cite{MWAA} is a managed orchestration service offered by AWS that simplifies the deployment and operation of Airflow environments in the cloud. MWAA handles infrastructure tasks such as provisioning, patching, scaling, and securing the environment. Google Cloud Composer \cite{GoogleCloudComposer} is GCP’s equivalent—a fully managed orchestration service built on Apache Airflow, providing similar capabilities to MWAA. Microsoft Azure offers Azure Data Factory \cite{AzureDataFactory}, a fully managed, serverless data integration and workflow orchestration service that enables users to build and schedule complex data pipelines for automating data movement and transformation. While Azure Data Factory provides a visual interface for building its native pipelines, it also includes a Workflow Orchestration Manager, which offers a managed Apache Airflow environment for defining DAGs programmatically in Python.

In addition, the three major cloud providers offer managed workflow orchestration services specifically designed to automate and manage Machine Learning (ML) workflows. Amazon SageMaker Pipelines \cite{AWSSageMaker} is a serverless orchestration service built for MLOps (Machine Learning Operations) and LLMOps (Large Language Model Operations), allowing workflows to be defined as DAGs. Its counterpart in GCP is Vertex AI Pipelines \cite{GoogleVertexAI}, which provides a fully managed, serverless platform for automating, monitoring, and governing ML workflows, supporting both Kubeflow Pipelines and TensorFlow Extended (TFX). Azure Machine Learning Pipelines \cite{AzureMLPipelines} offers similar capabilities, enabling ML workflows to run on serverless infrastructure, managed compute instances, or dedicated compute clusters.

For special purpose commercial brokers, we can mention DNAnexus \cite{DNAnexus}, which is one of the most complete commercial examples of WaaS brokers for managing, analyzing, and sharing large-scale genomic and biomedical data. It provides a secure, scalable environment for running scientific workflows and pipelines. It supports many workflow languages and tools such as CWL, WDL, and Nextflow, and executes the workflows on AWS and Azure Cloud platforms. It has different pricing models for enterprises, based on the compute and storage usages, the number of active users, and the overhead of running each workflow. The Seven Bridges Platform \cite{SevenBridges} is a similar commercial platform for running genomics and bioinformatic workflows (defined using CWL or WDL) on AWS, GCP and Microsoft Azure platforms. 

Hybrid quantum/classical scientific workflows are supported by all major cloud providers. Amazon offers AWS Braket Hybrid Jobs \cite{AWSBraketHybrid}, which enables the execution of hybrid quantum-classical algorithms—such as Variational Quantum Algorithms (VQAs) and Quantum Machine Learning (QML)—by combining classical AWS resources with Quantum Processing Units (QPUs). However, AWS receives the program as a Python script rather than an explicit Directed Acyclic Graph (DAG), although the internal execution can conceptually be represented as a DAG. Other Quantum Computing as a Service (QCaaS) providers follow a similar approach. For instance, IBM supports hybrid workflows through Qiskit \cite{Qiskit}, Google through Cirq \cite{Cirq}, and Microsoft Azure via its Quantum Development Kit (QDK) and Q\# language \cite{QSharp}, in addition to supporting Qiskit and Cirq.

In contrast, Covalent \cite{Covalent} is a hybrid quantum/classical WaaS broker that explicitly uses DAGs for workflow definition. It allows users to define workflows through Python decorators, enabling a more transparent and modular approach to hybrid workflow composition. Covalent is available in both a free open-source version and a commercial version, which includes pay-as-you-go pricing and enterprise contract-based plans.

\end{document}